\documentclass[aps,pre,showpacs,superscriptaddress,twocolumn]{revtex4}

\usepackage{amsmath}
\usepackage{amssymb}
\usepackage{graphicx}
\usepackage{color}
\usepackage{subfigure}

\def\vec#1{{\bf #1}}
\def\bold#1{{\boldsymbol #1}}

\begin{document}
\title{Bilayer-thickness-mediated interactions between integral membrane proteins}
\author{Osman Kahraman}
\affiliation{Department of Physics \& Astronomy and Molecular and Computational Biology Program, Department of Biological Sciences, University of Southern California, Los Angeles, CA 90089, USA}
\author{Peter D. Koch}
\affiliation{Department of Systems Biology, Harvard Medical School, Boston, MA 02115, USA}
\author{William S. Klug}
\affiliation{Department of Mechanical and Aerospace Engineering, University of California, Los Angeles, CA 90095, USA}
\author{Christoph A. Haselwandter}
\affiliation{Department of Physics \& Astronomy and Molecular and Computational Biology Program, Department of Biological Sciences, University of Southern California, Los Angeles, CA 90089, USA}

\begin{abstract}
Hydrophobic thickness mismatch between integral membrane proteins and the surrounding lipid bilayer can produce lipid bilayer thickness deformations. Experiment and theory have shown that protein-induced lipid bilayer thickness deformations can yield energetically favorable bilayer-mediated interactions between integral membrane proteins, and large-scale organization of integral membrane proteins into protein clusters in cell membranes. Within the continuum elasticity theory of membranes, the energy cost of protein-induced bilayer thickness deformations can be captured by
considering compression and expansion of the bilayer hydrophobic core, membrane tension, and bilayer bending, resulting in biharmonic equilibrium equations describing the shape of lipid bilayers for a given set of bilayer-protein boundary conditions. Here we develop a combined analytic and numerical methodology for the solution of the equilibrium elastic equations associated with protein-induced lipid bilayer deformations. Our methodology allows accurate prediction of thickness-mediated protein interactions for arbitrary protein symmetries at arbitrary protein separations and relative orientations. We provide exact analytic solutions for cylindrical integral membrane proteins with constant and varying hydrophobic thickness, and develop perturbative analytic solutions for non-cylindrical protein shapes. We complement these analytic solutions, and assess their accuracy, by developing both finite element and finite difference numerical solution schemes. We provide error estimates of our numerical solution schemes and systematically assess their convergence properties. Taken together, the work presented here puts into place an analytic and numerical framework which allows calculation of bilayer-mediated elastic interactions between integral membrane proteins for the complicated protein shapes suggested by structural biology and at the small protein separations most relevant for the crowded membrane environments provided by living cells.

\end{abstract}


\maketitle

\section{Introduction}
\label{secIntro}

In many cell types, cell membranes are composed \cite{phillips12} of a diverse array of lipids, organized as a lipid bilayer, and membrane proteins, which play a central role in most cellular processes. Membrane proteins are rigid compared to the surrounding lipid bilayer \cite{mouritsen93,engelman05,jensen04,andersen07}. Thus, the lipid bilayer typically deforms to accommodate membrane proteins and, in particular, the bilayer hydrophobic thickness is compressed or expanded compared to the preferred bilayer thickness in the absence of membrane proteins \cite{jensen04,lundbaek06,andersen07,phillips09,mcintosh06,brown12,engelman05,mouritsen93}. Distinct conformations of a membrane protein generally yield distinct
energy costs of protein-induced lipid bilayer deformations. As a result, the lipid bilayer can serve as a ``splint'' stabilizing certain protein conformations \cite{andersen07} and thereby regulate protein function~\cite{mouritsen93,jensen04,engelman05,andersen07,mcintosh06,phillips09,brown12,lundbaek06}.
In agreement with this general picture, experiments have revealed \cite{lundbaek10,greisen11,brohawn12,schmidt12,brohawn14,anishkin13,anishkin14,milescu09}
that, across the kingdoms of life, central biological functions of integral membrane proteins such as ion exchange and signaling are regulated by the mechanical properties of the surrounding lipid bilayer, with the hydrophobic regions of membrane proteins coupling to the hydrophobic regions of lipid bilayers \cite{mitra04,sonntag11,krepkiy09}. In particular, elastic bilayer thickness deformations have been found \cite{mouritsen93,jensen04,engelman05,andersen07,mcintosh06,phillips09,brown12,lundbaek06}
to regulate the functions of a diverse range of integral membrane~proteins.

Cell membranes are crowded with membrane proteins \cite{engelman05,takamori06,dupuy08,linden12}, with a typical mean center-to-center distance $d\approx10$~nm between neighboring proteins \cite{phillips09}. As a result, the elastic decay length of protein-induced lipid bilayer thickness deformations \cite{wiggins04,wiggins05,ursell08} is comparable to the typical edge-to-edge spacing of proteins in cell membranes
\cite{phillips09}, yielding thickness-mediated interactions between membrane proteins \cite{harroun99,grage11,goforth03,botelho06,phillips09}. For the small protein separations relevant for cell membranes, thickness-mediated interactions between integral membrane proteins can be $> 10$~$k_B T$ in magnitude \cite{ursell07,phillips09} and, depending on the hydrophobic thickness
of neighboring membrane proteins, be energetically favorable or unfavorable.

The lipid bilayer elasticity theory \cite{seifert97,boal02,safran03}
underlying the description of protein-induced bilayer deformations and bilayer-mediated
protein interactions has a rich and distinguished history, dating back to the classic work of W. Helfrich \cite{helfrich73}, P.~B. Canham \cite{canham70}, E.~A. Evans \cite{evans74}, and H.~W. Huang \cite{huang86}. According to this classic theory, membrane proteins may, in addition to thickness-mediated interactions \cite{dan93,dan94,aranda96,dan98,harroun99b,partenskii04,brannigan07,ursell07,CAH2013a,OWC1},
also interact \cite{fournier99,phillips09} via bilayer curvature deformations
\cite{goulian93,weikl98,kim98,kim00,muller05,muller05b,kim08,auth09,muller10,frese08,reynwar11,bahrami14,dommersnes99,evans03,weitz13,yolcu14}
and bilayer fluctuations \cite{goulian93,dommersnes99,evans03,weitz13,yolcu14,golestanian96,golestanian1996b,weikl01,lin11}.
While the competition between thickness-,\linebreak curvature-, and fluctuation-mediated
protein interactions depends on the properties of the specific lipids and
membrane proteins under consideration, one generally expects \cite{ursell07,phillips09} that thickness-mediated protein interactions are strong and short-ranged, and that curvature- and fluctuation-mediated protein interactions are weak and long-ranged.

The classic elasticity theory of protein-induced lipid bilayer deformations can be extended in various ways 
\cite{gil98,brannigan06,brannigan07,west09,may04,may99,bohinc03,may07,watson11,watson13,jablin14,rangamani14,bitbol12,partenskii02,partenskii03,partenskii04,kim12,yoo13,yoo13b,lee13}
to account for detailed molecular properties of lipids such as lipid tilt and lipid intrinsic curvature \cite{dan93,dan94,aranda96,dan98,fournier99}, yielding additional modes of bilayer-mediated protein interactions. Theoretical studies of bilayer-mediated protein interactions have largely
focused on idealized (often cylindrical or conical) protein shapes which do not correspond to any particular membrane protein, and proteins at large $d$. In contrast, bilayer-mediated protein interactions at small $d$ are most relevant for the crowded environment provided by cell membranes \cite{engelman05,takamori06,dupuy08,linden12},
while modern structural biology suggests a rich picture of membrane protein shape with experimental surveys of the protein content in various cell membranes \cite{takamori06,yun11,linden12} indicating great diversity in the oligomeric states and symmetries of membrane proteins.

The central goal of this article is to provide a detailed discussion of a combined analytic and numerical framework \cite{CAH2013a,OWC1,CAH2013b,OPWC1} which allows prediction of lipid bilayer-mediated elastic interactions between integral membrane proteins at arbitrary $d$ for the protein shapes suggested by structural studies. We focus here on protein-induced lipid bilayer thickness deformations, which have been found
\cite{mouritsen93,jensen04,engelman05,andersen07,mcintosh06,phillips09,brown12,lundbaek06,harroun99,goforth03,botelho06,grage11}
to play central roles in regulation of protein function and bilayer-mediated
protein interactions in a wide range of experimental systems 
\cite{dan93,dan94,aranda96,dan98,harroun99b,partenskii04,brannigan07,ursell07,huang86,helfrich90,nielsen98,nielsen00,harroun99,partenskii02,partenskii03,kim12,lundbaek10,greisen11,wiggins04,wiggins05,ursell08,grage11,CAH2013a,CAH2013b,OWC1,OPWC1,mondal11,mondal12,mondal13,mondal14,CAH2014a}.
Using this mathematical framework we have shown previously that the shape of integral membrane proteins, and resulting structure of lipid bilayer thickness deformations, can play a crucial role in the regulation of protein function by lipid bilayers \cite{OWC1,CAH2013b}, and that bilayer thickness-mediated interactions between integral membrane proteins can be strongly directional and dependent on protein shape \cite{CAH2013a,OWC1,OPWC1,CAH2014a}. Thus, in addition to the magnitude of the bilayer-protein hydrophobic mismatch
\cite{engelman05,mouritsen93,jensen04,mcintosh06,lundbaek06,andersen07,phillips09,brown12,harroun99,goforth03,botelho06,grage11},
protein shape may be a crucial determinant of membrane protein regulation
by lipid bilayers and bilayer-mediated protein interactions.

\begin{figure}[t!]
\includegraphics[width=0.92\columnwidth]{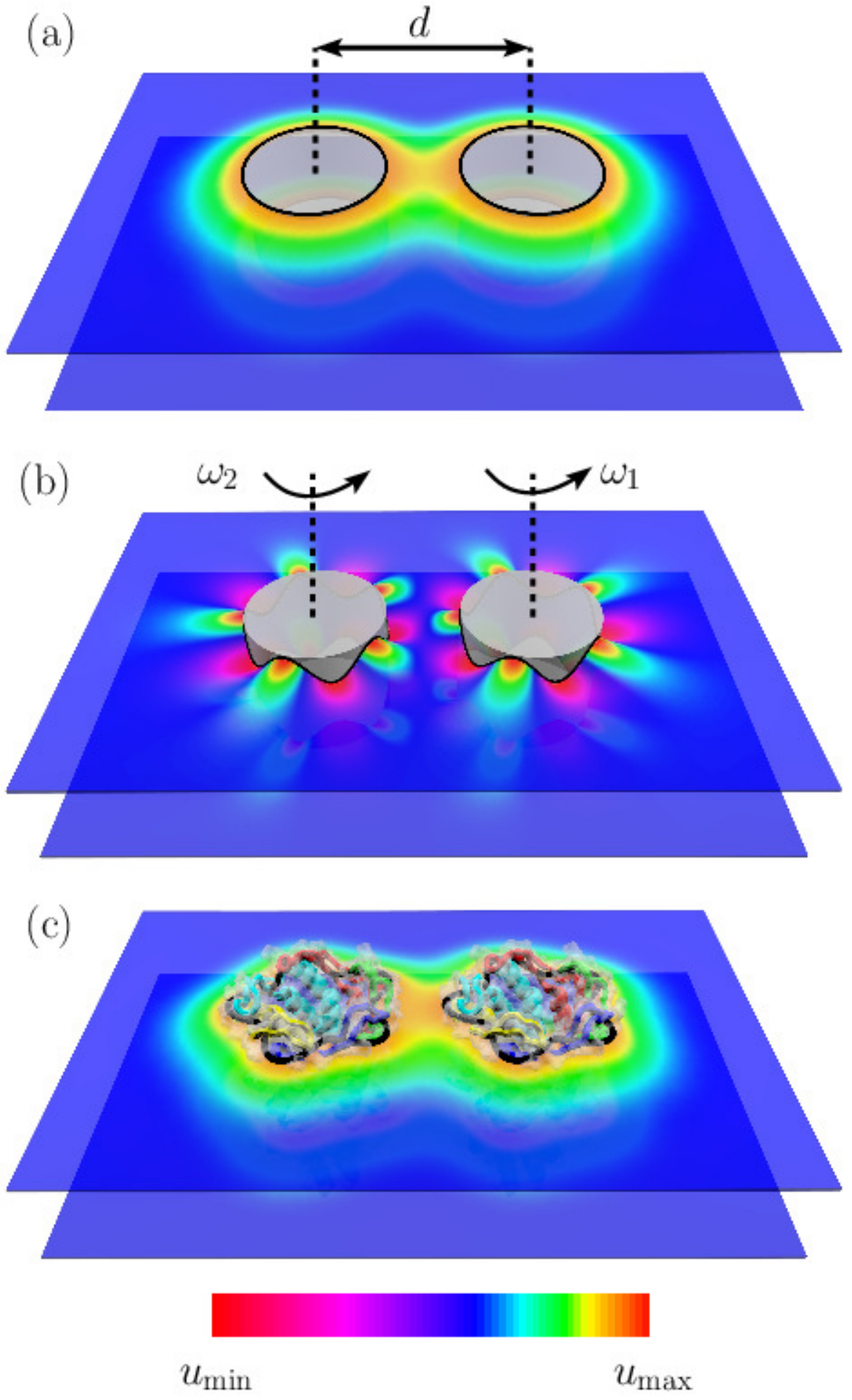}
\caption{(Color online) Overlapping lipid bilayer thickness deformation fields can yield bilayer thickness-mediated interactions between membrane proteins. Bilayer thickness deformations $u$ due to interacting membrane proteins obtained using (a) the cylinder model, (b) the crown model, and (c) the clover-leaf model of integral membrane proteins (see Sec.~\ref{secElasticModel} for further details). The thickness deformations in panels (a) and (b) were obtained
through exact analytic minimization of the thickness deformation energy (see Sec.~\ref{secAnalyticSol}), and the thickness deformations in panel (c) were calculated using finite elements (see Sec.~\ref{secFE}).
The clover-leaf shape in panel (c) provides a simple coarse-grained model \cite{CAH2013a,CAH2013b,OWC1,OPWC1} of the observed closed-state structure of the pentameric mechanosensitive channel of large conductance \cite{chang98} (protein structural data shown as ribbon diagrams; Protein Data Bank accession number 2OAR). The calculated lipid bilayer thickness deformations depend on lipid and protein properties, the protein center-to-center distance $d$, and, for the crown and clover-leaf models, the protein orientations $\omega_{1,2}$. The color scale ranges from $u_{\text{min}}=-0.6$~nm to $u_{\text{max}}=0.4$~nm.
}
 \label{figIllust}
\end{figure}

We develop, illustrate, and test our analytic and numerical framework
for calculating bilayer-mediated protein interactions using the protein shapes shown in Fig.~\ref{figIllust}, which embody key mechanisms by which protein crowding and protein shape may affect protein-induced lipid
bilayer deformations. The most straightforward model of protein-induced lipid bilayer thickness deformations assumes a circular protein
cross section with constant boundary conditions along the bilayer-protein
interface [see Fig.~\ref{figIllust}(a)]. The resulting ``cylinder model'' of membrane proteins allows investigation
of thickness-mediated protein interactions in crowded membranes without the further complications introduced by a complicated protein shape. The cylinder model has been used before in a number of different settings 
\cite{huang86,wiggins05,ursell08,helfrich90,andersen07,phillips09,nielsen98,nielsen00}
to describe protein-induced bilayer thickness deformations.

Aside from interactions with neighboring membrane proteins, angular variations
in the bilayer-protein boundary conditions \cite{mondal11,mondal12,mondal13,mondal14,CAH2013a}
or a non-circular protein cross
section \cite{CAH2013a,CAH2013b,OWC1,OPWC1,CAH2014a} may also break rotational symmetry of bilayer thickness deformations about the protein center. Simple representations of these two features of protein shape are provided \cite{CAH2013a,CAH2013b,OWC1,OPWC1} by the ``crown model'' [see Fig.~\ref{figIllust}(b)], in which we allow for angular variations of the protein hydrophobic thickness while assuming a circular protein cross section, and by the ``clover-leaf model'' [see Fig.~\ref{figIllust}(c)], in which we assume constant boundary conditions along the bilayer-protein interface but allow for a non-circular shape of the protein cross section. In general, membrane proteins will have both a non-circular cross section and variable hydrophobic thickness. The crown and clover-leaf models allow us to isolate these two possible origins of angular anisotropy of protein-induced
lipid bilayer deformations. An important difference between the cylinder model, and the crown and clover-leaf models, is that for the latter models bilayer-mediated protein interactions are inherently directional and depend not only on the protein separation $d$ but also on the protein orientations $\omega_i$, where the index $i=1,2,\dots$ denotes different membrane proteins [Figs.~\ref{figIllust}(b,c)].

The organization of this article is as follows. Section~\ref{secElasticModel}
provides a detailed discussion of the elastic energy of lipid bilayer thickness
deformations, and the cylinder, crown, and clover-leaf models of integral membrane proteins. In Sec.~\ref{secAnalyticSol} we obtain, based on Refs.~\cite{huang86,dan93,dan94,aranda96,dan98,goulian93,weikl98}, analytic solutions of the thickness deformation fields and thickness deformation energies due
to cylinder, crown, and clover-leaf shapes at arbitrary $d$ and protein orientations. These analytic
solutions are exact for cylinder and crown shapes, and perturbative for clover-leaf shapes. As described in Sec.~\ref{secNumericalSol}, we complement our analytic solutions, and assess their validity, by developing numerical solution schemes based on finite element (FE) and finite difference (FD) solution procedures. In particular, the FE approach described here offers a straightforward way of representing the complicated protein shapes suggested by membrane structural biology, and is efficient and accurate enough to enable prediction of the directional thickness-mediated interactions of hundreds of integral membrane proteins at arbitrary $d$ and protein orientations \cite{OWC1,OPWC1} with reasonably coarse computational grids. In Sec.~\ref{secCylinder2} we provide a detailed comparison between analytic and numerical results for the thickness deformation fields and thickness deformation energies implied by the cylinder model in the non-interacting and interacting regimes of $d$. Sections~\ref{secCrownResults} and~\ref{secCloverResults} provide similar comparisons between analytic and numerical results for the crown and clover-leaf models. A summary of our results and conclusions can be found in Sec.~\ref{secSummary}.

\section{Elastic model of protein-induced bilayer thickness deformations}
\label{secElasticModel}

\subsection{Elastic thickness deformation energy}

In the standard elasticity theory of bilayer-protein interactions 
\cite{jensen04,lundbaek06,andersen07,phillips09}, integral membrane proteins are assumed to be rigid membrane inclusions which
deform the surrounding lipid bilayer \cite{boal02,safran03,seifert97}. In
the simplest formulation, bilayer deformations can then be captured by two coupled scalar fields $h_+$ and $h_-$ which define the positions of the hydrophilic-hydrophobic interface in the outer and inner lipid bilayer leaflets, respectively. It is mathematically convenient to express $h_+$ and $h_-$ in terms of the midplane deformation
field
\begin{equation}
h=\frac{1}{2} \left(h_+ + h_- \right)
\end{equation}
and the thickness deformation field
\begin{equation}
u=\frac{1}{2} \left(h_+ - h_- -2 a\right)\,,
\end{equation}
in which $a$ is one-half the unperturbed hydrophobic thickness of the lipid bilayer.

To leading order, the elastic energies governing $h$ and $u$ decouple from each other \cite{huang86,fournier99}. In the most straightforward model of bilayer-protein interactions 
\cite{huang86,goulian93,dan93,weikl98,jensen04,lundbaek06,andersen07,phillips09,boal02,safran03,seifert97}, the energy cost of midplane deformations is then captured by the Helfrich-Canham-Evans energy \cite{canham70,helfrich73,evans74} and, within the Monge representation, the energy cost of thickness deformations is of the form~\cite{huang86,andersen07,ursell08}
\begin{equation} \label{energy}
{\textstyle G=\frac{1}{2}}\int dx dy {\textstyle\left\{K_b (\nabla^2 u)^2+K_t \left(\frac{u}{a}\right)^2+\tau
\left[2 \frac{u}{a}+(\nabla u)^2 \right] \right\}}\,,
\end{equation}
where $K_b$ is the bending rigidity of the lipid bilayer, $K_t$ is the stiffness associated with thickness deformations, and $\tau$ is the membrane tension. The effective parameters $K_b$, $K_t$, and $a$ in Eq.~(\ref{energy}) encapsulate bilayer material properties relevant for protein-induced bilayer thickness
deformations and depend on the bilayer composition \cite{rawicz00,rawicz08,nagle13}. Typical values measured in experiments are $K_b = 20$ $k_B T$, $K_t = 60$ $k_B T/$nm$^2$, and $a=1.6$~nm \cite{andersen07,phillips09}, which we use for all the numerical calculations described here.

The classic model of protein-induced lipid bilayer thickness deformations in Eq.~(\ref{energy}) employs the Monge representation of surfaces, $u=u(x,y)$, with Cartesian coordinates $(x,y)$, and only considers leading-order terms
in $u$ and its derivatives. The former assumption can be justified by noting that thickness deformations generally decay rapidly compared to midplane deformations, with typical thickness and midplane decay lengths $\approx 1$~nm and $\approx 5$--500~nm \cite{phillips09,ursell08}, respectively. The validity of the latter assumption depends on the specific properties of the lipid bilayer and protein under consideration, but for experimental model systems \cite{nielsen98,nielsen00,wiggins04,wiggins05,ursell07,ursell08,grage11,OPWC1} one typically finds $u/a<0.3$ and $\|\nabla u \| < 0.2$. Hence, bilayer overhangs and higher-order corrections to Eq.~(\ref{energy}) can often be neglected when describing protein-induced lipid bilayer thickness deformations, with the bilayer midplane being approximately parallel to the reference plane invoked in the Monge representation of surfaces.

The terms $K_b \left(\nabla^2 u \right)^2$ and $K_t \left(u/a\right)^2$ in Eq.~(\ref{energy}) provide lowest-order descriptions of the energy cost of bilayer bending, and compression and expansion, of the bilayer hydrophobic
core, respectively. For generality we allow for the two tension terms $2 \tau u/a$ and $\tau \left(\nabla u\right)^2$ in Eq.~(\ref{energy}), which account \cite{boal02,safran03,seifert97} for stretching deformations tangential to the leaflet surfaces and changes in the projection of the bilayer area onto the reference plane, respectively. The minimal model in Eq.~(\ref{energy}) can be extended in a variety of ways to account for more detailed properties of lipid bilayers including lipid tilt \cite{fournier99,may07,watson11,watson13,jablin14,rangamani14},
lipid intrinsic curvature \cite{dan93,dan94,aranda96,dan98,brannigan06,brannigan07,west09}, inhomogeneous deformation of lipid volume and effects of Gaussian curvature on protein-induced bilayer deformations \cite{brannigan06,brannigan07,west09}, asymmetric bilayer thickness deformations \cite{brannigan06,brannigan07,west09,bitbol12}, and protein-induced local modulation of bilayer elastic properties \cite{partenskii02,partenskii03,partenskii04,kim12,yoo13,yoo13b,lee13}.

The lipid bilayer thickness deformation energy in Eq.~(\ref{energy}) provides a simple model of thickness-mediated protein interactions, as well as the coupling between protein function and bilayer thickness deformations, and has the appealing property that all the material parameters entering Eq.~(\ref{energy}) can be measured directly in experiments. For given bilayer-protein boundary conditions (see Sec.~\ref{secShape}), minimization of Eq.~(\ref{energy}) completely specifies the lowest-energy bilayer thickness configuration and its associated energy cost. Models based on Eq.~(\ref{energy}) have been found to capture the basic experimental phenomenology of bilayer-protein interactions for
gramicidin channels 
\cite{huang86,helfrich90,nielsen98,nielsen00,harroun99,harroun99b,partenskii02,partenskii03,partenskii04,kim12,lundbaek10,greisen11}, the mechanosensitive channel of large conductance (MscL) 
\cite{wiggins04,wiggins05,ursell07,ursell08,grage11,CAH2013a,CAH2013b,OWC1,OPWC1}, G-protein coupled receptors \cite{mondal11,mondal12,mondal13,mondal14}, the bacterial leucine transporter \cite{mondal14}, and chemoreceptor lattices \cite{CAH2014a}, as well as a variety of other integral membrane proteins \cite{andersen07,phillips09,jensen04,lundbaek06,mcintosh06,brown12}. While we focus here on elastic thickness deformations, midplane deformations may
generally also contribute \cite{goulian93,weikl98,wiggins05,phillips09} to bilayer-mediated protein interactions, and the regulation of protein function by bilayer mechanical properties.

Minimization of Eq.~(\ref{energy}) can be performed by solving the appropriate
Euler-Lagrange equation, which is given by
\begin{equation} \label{genBihpreu}
K_b \nabla^4 u - \tau \nabla^2 u+\frac{K_t}{a^2} u+\frac{\tau}{a}=0\,.
\end{equation}
The analytic solution of Eq.~(\ref{genBihpreu}) is facilitated \cite{CAH2013b} by introducing the function
\begin{equation} \label{transfu}
\bar u(x,y)= u(x,y)+\frac{\tau a}{K_t}\,,
\end{equation}
in terms of which Eq.~(\ref{genBihpreu}) can be expressed as
\begin{equation} \label{genBih}
\left(\nabla^2 - \nu_+\right) \left(\nabla^2 - \nu_-\right) \bar u=0\,,
\end{equation}
where  
\begin{equation} \label{defnu}
\nu_\pm = \frac{1}{2 K_b} \left[\tau \pm \left(\tau^2-\frac{4 K_b K_t}{a^2} \right)^{1/2} \right]\,.
\end{equation}

To analytically calculate the thickness deformation energy associated with
the solutions of Eq.~(\ref{genBihpreu}) we use Eqs.~(\ref{genBihpreu}) and~(\ref{transfu}) to rearrange the thickness deformation energy in Eq.~(\ref{energy}) as
\begin{eqnarray} \nonumber
G&=&G_1+\\&&\frac{1}{2}}\int dx dy {\textstyle \nabla \cdot \left[K_b (\nabla \bar u) \nabla^2 \bar u-K_b \bar u  \nabla^3 \bar u+\tau \bar u \nabla \bar u \right]\,, \nonumber\\&&\label{energy2}
\end{eqnarray}
where the term
\begin{equation}\label{energyG1}
G_1=-\frac{1}{2} \int dx dy \frac{\tau^2}{K_t}
\end{equation}
is independent of $\bar u$ and arises, for $\tau > 0$, due to relaxation of the ``loading device'' producing membrane tension \cite{ursell08} via uniform compression of the bilayer hydrophobic core. Since $G_1$ does not contribute to the energy cost of protein-induced bilayer deformations we subtract $G_1$ from $G$, $G \to G-G_1$. Using Gauss's theorem, we then find
\begin{equation} \label{EvalEnergyLine}
G=\frac{1}{2}\int dl \, {\textstyle \mathbf{\hat n} \cdot \left[K_b (\nabla \bar u) \nabla^2 \bar u-K_b \bar u  \nabla^3 \bar u+\tau \bar u \nabla \bar u \right]}\,,
\end{equation}
where the line integrals $\int dl$ are to be taken along all bilayer-protein
interfaces, with the bilayer unit normal vectors $\mathbf{\hat n}$ perpendicular to the bilayer-protein interfaces and pointing towards the proteins, and
we assume \cite{huang86,nielsen98,nielsen00} that $\bar u$, as well as its derivatives, go to zero far from the proteins.

\subsection{Modeling protein shape}
\label{secShape}

Following previous work on lipid bilayer-protein interactions 
\cite{huang86,goulian93,weikl98,dan93,jensen04,lundbaek06,andersen07,phillips09}
we model integral membrane proteins as rigid membrane inclusions of fixed shape and hydrophobic thickness. The specific properties of a given membrane protein enter our description of bilayer-protein interactions through the shape of the protein cross section, and through the boundary conditions on $u$ along the bilayer-protein interface. As described
in Sec.~\ref{secIntro}, we consider here three distinct models of protein shape: the cylinder model, the crown model, and the clover-leaf model. These minimal models do not provide detailed descriptions of protein shape but, rather, aim to encapsulate the features of a given protein hydrophobic surface most crucial for protein-induced lipid bilayer thickness deformations. As noted above, midplane deformations decouple to leading order from thickness deformations. Hence, the models of protein-induced bilayer
thickness deformations considered here can be easily complemented by corresponding models of protein-induced bilayer midplane deformations \cite{goulian93,weikl98,wiggins05,phillips09}, which capture separate aspects of bilayer-protein interactions.

\subsubsection{Cylinder model}
\label{secCylinder}

A straightforward description of the effect of protein shape on protein-induced
lipid bilayer thickness deformations is provided by the cylinder model of integral membrane proteins [Fig.~\ref{figIllust}(a)] 
\cite{huang86,wiggins05,ursell08,helfrich90,andersen07,phillips09,nielsen98,nielsen00}. Introducing the polar coordinates $(r_i,\theta_i)$ with
the center of membrane protein $i$ as the origin, the boundary conditions for protein $i$ can be written as
\begin{eqnarray} \label{bc1f}
u(r_i,\theta_i)\big|_{r_i=C_i(\theta_i)}&=&U_i\,,\\ \label{bc2f}
\mathbf{\hat n} \cdot \nabla u(r_i,\theta_i)\big|_{r_i=C_i(\theta_i)}&=&U_i^\prime\,,
\end{eqnarray}
where the boundary curve $C_i(\theta_i)=R_i$ for a membrane protein
with a circular cross section of radius $R_i$, and the constants \cite{wiggins05}
\begin{eqnarray} \label{bc1}
U_i &=& \frac{1}{2} \left(W_i - 2a\right)\,,\\ \label{bc2}
U_i^\prime &=& \frac{1}{2} \left(H_+^\prime-H_-^\prime\right)\,,
\end{eqnarray}
where $W_i$ is the hydrophobic thickness of protein $i$ and the $H_\pm^\prime$ correspond to the normal derivatives of $h_\pm$ evaluated along the bilayer-protein boundary.

Equation~(\ref{bc1}) assumes perfect hydrophobic matching between the membrane protein and the lipid bilayer \cite{harroun99,harroun99b,andersen07,phillips09}. This assumption is expected to break down for a large enough hydrophobic mismatch between the protein and the (undeformed) lipid bilayer \cite{nielsen98,nielsen00,mondal11,mondal12,mondal13,mondal14},
in which case $W_i$ corresponds to the effective hydrophobic thickness of
the membrane protein. Unless indicated otherwise, we use $W_i=3.8$~nm for the numerical calculations described here, which approximately corresponds to \cite{ursell08,elmore03} the hydrophobic thickness of the observed structure of closed pentameric MscL \cite{chang98}. Following previous work on MscL-induced lipid bilayer thickness deformations employing the cylinder model of integral membrane proteins \cite{wiggins04,wiggins05,ursell07,ursell08,phillips09}, we use $R_i=2.3$~nm, which yields an area of the transmembrane protein cross section consistent with the observed structure of closed pentameric MscL
\cite{chang98}. A number of different choices for the boundary condition in Eq.~(\ref{bc2f}) have been investigated 
\cite{huang86,helfrich90,nielsen98,nielsen00,harroun99,harroun99b,partenskii02,partenskii03,partenskii04,bitbol12,kim12,lee13,brannigan06,brannigan07,west09}.
In particular, $U_i^\prime$ may be chosen based on experimental observations
or molecular dynamics simulations, or may be regarded as a free parameter to be fixed as part of the energy minimization procedure. We follow here previous theoretical work on the experimental phenomenology of gramicidin channels \cite{huang86,harroun99,harroun99b} and MscL \cite{wiggins04,wiggins05,ursell07,ursell08,phillips09} which suggests that, to a first approximation, $U_i^\prime=0$.

\subsubsection{Crown model}
\label{secCrown}

The hydrophobic thickness of integral membrane proteins is generally expected to vary along the bilayer-protein interface \cite{sonntag11,krepkiy09,mondal11,mondal12,mondal13,mondal14},
yielding anisotropic bilayer thickness deformations and, in the case of two or more membrane proteins in sufficiently close proximity, directional interactions \cite{CAH2013a}. To study the generic effects of a variable protein hydrophobic thickness on bilayer-mediated protein interactions we replace the constant $U_i$ in Eq.~(\ref{bc1f}) by \cite{CAH2013a}
\begin{equation} \label{VarU}
U_{i}(\theta_i)=U_i^0+ \delta_i \cos s \left(\theta_i-\omega_i\right)\,,
\end{equation}
where $U_i^0$ is the average hydrophobic mismatch, $\delta_i$ is the
magnitude of mismatch modulations, $s$ is the protein symmetry, and $\omega_i$ parametrizes the orientation of protein $i$. For each bilayer leaflet, Eq.~(\ref{VarU})
yields a periodic modulation of the protein hydrophobic surface which resembles the shape of a crown [Fig.~\ref{figIllust}(b)], and we therefore refer to Eq.~(\ref{VarU}) as the crown model of integral membrane proteins.

For our numerical calculations we use the values $U_i^0=-0.1$~nm, $\delta_i=0.5$~nm, and $s=5$ in Eq.~(\ref{VarU}), and vary $\omega_i$ to explore bilayer thickness-mediated interactions for a range of relative protein orientations. We choose all other parameter values as described for the cylinder model of membrane proteins.
Even for non-interacting membrane proteins, this parametrization of the
crown model in Eq.~(\ref{VarU}) yields a maximum magnitude of the gradient of bilayer thickness deformations $\approx 1$, and therefore produces thickness
deformations which lie at the limit of applicability of the leading-order energy in Eq.~(\ref{energy}). Furthermore, for interacting membrane proteins we generally find with this
parametrization of the crown model that the maximum magnitude of the gradient of bilayer thickness deformations $>1$. As a result, the numerical estimates of the thickness deformation energy obtained for this parametrization of Eq.~(\ref{VarU}) are of limited physical significance. We allow here for such large magnitudes of the gradient of bilayer thickness deformations in order to explore the mathematical limits of applicability of our analytic and numerical solution
procedures (see Sec.~\ref{secCrownResults}).

\subsubsection{Clover-leaf model}
\label{secClover}

Membrane structural biology has produced a rich and diverse picture of membrane
protein shape, which suggests \cite{spencer02,takamori06,yun11,linden12} that integral membrane proteins can occur in a variety of different oligomeric states and transmembrane shapes. Distinct oligomeric states of membrane proteins generally yield distinct symmetries of the protein cross section which, in turn, induce distinct symmetries of lipid bilayer thickness deformations
\cite{mondal11,mondal12,mondal13,mondal14,CAH2013a,CAH2013b,OWC1,OPWC1}.
The resulting non-trivial structure of bilayer thickness deformations can yield substantial deviations from the energy cost of protein-induced
bilayer thickness deformations implied by the cylinder model of integral membrane proteins. In particular, for MscL it has been found \cite{CAH2013a,CAH2013b,OWC1,OPWC1}
that the elastic energy of protein-induced bilayer thickness deformations provides
a signature of the protein oligomeric state, with distinct MscL oligomeric states yielding distinct MscL gating characteristics and directional bilayer-mediated
protein interactions.

A simple coarse-grained model of the cross sections of a diverse range of membrane proteins is provided by the clover-leaf model \cite{CAH2013a,CAH2013b,OWC1}
\begin{equation} \label{boundCclover}
C_{i}(\theta_i)=R_i \left[1+\epsilon_i \cos s \left(\theta_i-\omega_i\right) \right]\,,
\end{equation}
where $\epsilon_i$ parametrizes the magnitude of the deviation of the protein cross section from the circle [Fig.~\ref{figIllust}(c)]. In particular, the structure of pentameric MscL observed in \textit{Mycobacterium tubercolosis} \cite{chang98} suggests \cite{CAH2013a,CAH2013b,OWC1} $s=5$, $R_i=2.27$~nm, and $\epsilon_i=0.22$, which we use for all the numerical calculations involving clover-leaf shapes described here. In general, the hydrophobic thickness
of integral membrane proteins is expected to vary along the boundaries of clover-leaf shapes. However, in order to isolate the effect of anisotropy in protein shape on bilayer thickness-mediated protein interactions we focus here on the simpler scenario of a constant hydrophobic thickness, and use for the clover-leaf model of integral membrane proteins the same boundary conditions on $u$ along the bilayer-protein interface as for the cylinder model [see Eqs.~(\ref{bc1f}) and~(\ref{bc2f})].

\section{Analytic solution}
\label{secAnalyticSol}

\begin{figure}[t!]
\includegraphics[width=\columnwidth]{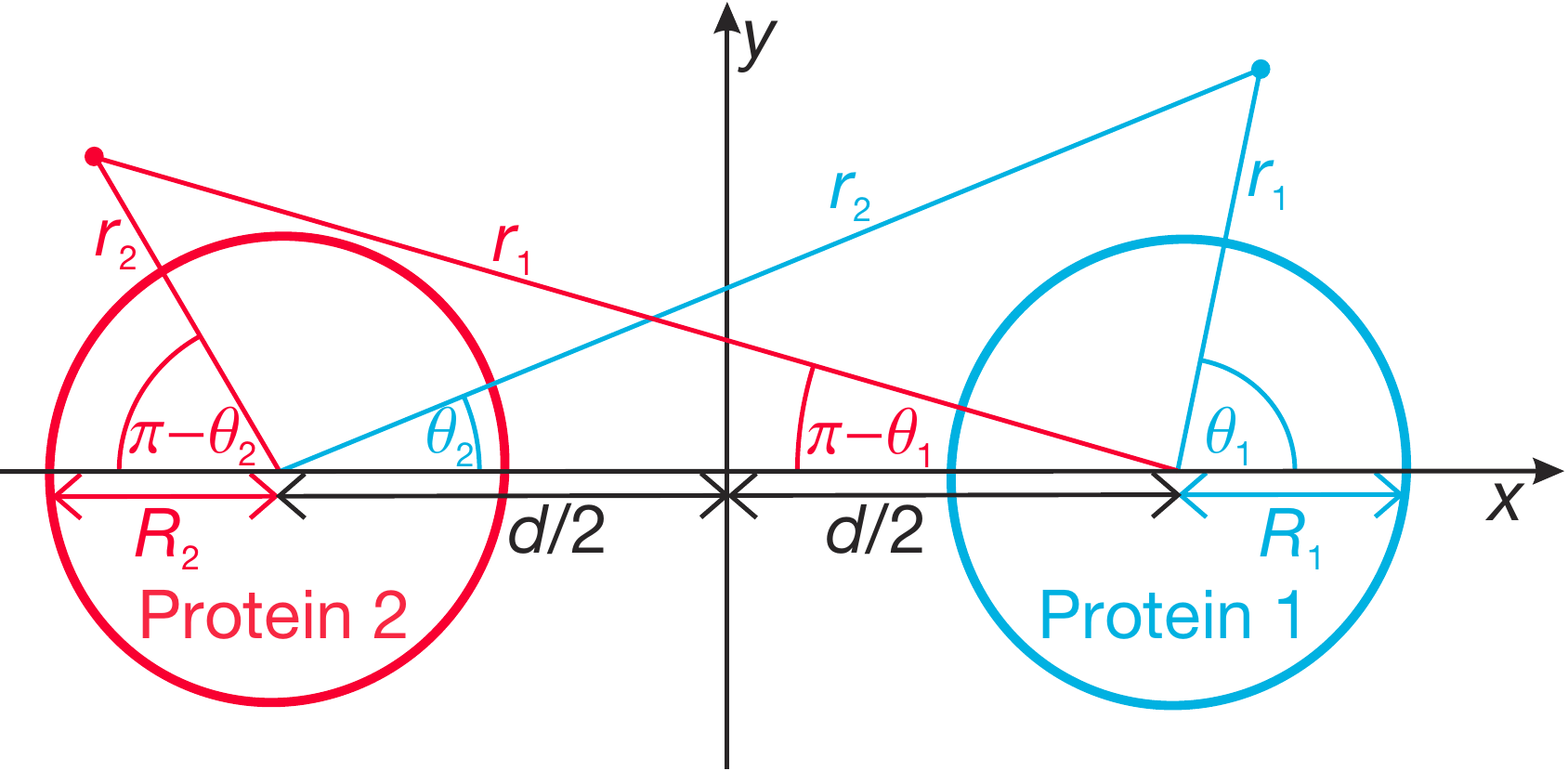}
\caption{(Color online) Two-center bipolar coordinate system for two membrane
 proteins with circular cross sections of radii $R_{1,2}$ separated by a center-to-center distance $d$ along the $x$-axis. The $x$-$y$ plane corresponds to the reference
plane used in the Monge representation of surfaces. Expressions for $(r_1,\theta_1)$ in terms of $(r_2,\theta_2)$ can be obtained by considering the coordinates of the red (left) point, and expressions for $(r_2,\theta_2)$ in terms of $(r_1,\theta_1)$ can be obtained by considering the coordinates of the blue (right)~point.}
\label{figBiPol}
\end{figure}

Building on earlier work on the lipid bilayer thickness deformations induced by
cylindrical membrane proteins \cite{huang86,dan93,dan94,aranda96,dan98} and bilayer curvature-mediated interactions between conical
membrane proteins in the far-field limit \cite{goulian93,weikl98}, we develop in this section analytic solutions \cite{CAH2013a,CAH2013b} of the bilayer
thickness-mediated interactions between integral membrane proteins at arbitrary protein separations
and relative orientations. To solve for the thickness-mediated interactions implied by Eq.~(\ref{energy})
for two membrane proteins, we employ a two-center
bipolar coordinate system (see Fig.~\ref{figBiPol}). For the sake of simplicity, we assume in Fig.~\ref{figBiPol} that the two proteins have circular cross sections with radii $R_{1,2}$. We relax this assumption below to capture interactions
between clover-leaf shapes. To mathematically relate the polar coordinates $(r_{1,2},\theta_{1,2})$
centered about proteins 1 and 2, we note from Fig.~\ref{figBiPol} the bipolar coordinate transformations
\begin{equation}
r_2=\left(d^2+r_1^2+2 d r_1 \cos \theta_1 \right)^{1/2}\,,
\end{equation}
$\cos \theta_2 = \left(d+r_1 \cos \theta_1\right)/r_2$, and $\sin \theta_2 = \left(r_1 \sin \theta_1\right)/r_2$. The corresponding transformations
for $r_1$ and $\sin \theta_1$ are symmetric in the protein indices, but $\cos \theta_1 = - \left(d-r_2 \cos \theta_2\right)/r_1$.

We solve the Euler-Lagrange equation~(\ref{genBih}) by making the ansatz \cite{weikl98,CAH2013a}
\begin{equation} \label{genSol}
\bar u=\bar u_1(r_1,\theta_1)+\bar u_2(r_2,\theta_2)\,,
\end{equation}
where the $\bar u_i(r_i,\theta_i)$ are the solutions of Eq.~(\ref{genBih}) for a single protein $i=1,2$, which are of the form \cite{huang86,zauderer83}
\begin{equation} \label{constructGenSol}
\bar u_i(r_i,\theta_i)=f_i^+(r_i,\theta_i) + f_i^-(r_i,\theta_i)\,,
\end{equation}
where the $f_i^\pm$ are solutions of the Helmholtz equations
\begin{equation} 
\nabla^2 f_i^\pm = \nu_{\pm} f_i^\pm\,.
\end{equation}
For the exterior of a circle of radius $R_i$, the above Helmholtz equations
are readily solved by separation of variables \cite{boas83,zauderer83}. Thus, the general single-protein solution of Eq.~(\ref{genBih}) can be constructed
from
\begin{eqnarray}  \nonumber
f_i^\pm(r_i,\theta_i) &=& A_{i,0}^\pm K_0(\sqrt{\nu_\pm} r_i)+ C_{i,0}^\pm I_0(\sqrt{\nu_\pm} r_i) \\ && \label{genSolSingle}
+\sum_{n=1}^\infty \left\{\mathcal{A}_{i,n}^\pm + \mathcal{B}_{i,n}^\pm+ \mathcal{C}_{i,n}^\pm+
\mathcal{D}_{i,n}^\pm\right\}  \,, \quad
\end{eqnarray}
where $A_{i,0}^\pm$ and $B_{i,0}^\pm$ are constants, the Fourier-Bessel terms
\begin{eqnarray}
\mathcal{A}_{i,n}^\pm&=& A_{i,n}^\pm K_n(\sqrt{\nu_\pm} r_i) \cos n \theta_i\,,\\
\mathcal{B}_{i,n}^\pm&=& B_{i,n}^\pm K_n(\sqrt{\nu_\pm} r_i) \sin n \theta_i\,,\\
\mathcal{C}_{i,n}^\pm&=& C_{i,n}^\pm I_n(\sqrt{\nu_\pm} r_i) \cos n \theta_i\,,\\
\mathcal{D}_{i,n}^\pm&=& D_{i,n}^\pm I_n(\sqrt{\nu_\pm} r_i) \sin n \theta_i\,,
\end{eqnarray}
the $I_j$ and $K_j$ with $j \geq 0$ are the modified Bessel functions of the first and second kind, and $A_{i,n}^\pm$, $B_{i,n}^\pm$, $C_{i,n}^\pm$, $D_{i,n}^\pm$ with $n \geq 1$ are constants. Assuming that $\bar u \to 0$ as $r_i \to \infty$ \cite{huang86,nielsen98,nielsen00}, we have $C_{i,0}^\pm=C_{i,n}^\pm = D_{i,n}^\pm =0$ for $n \geq 1$, and Eq.~(\ref{genSolSingle}) reduces to \cite{CAH2013b}
\begin{equation} \label{genSolSingle2}
f_i^\pm(r_i,\theta_i) = A_{i,0}^\pm K_0(\sqrt{\nu_\pm} r_i) 
+\sum_{n=1}^N \left\{\mathcal{A}_{i,n}^\pm + \mathcal{B}_{i,n}^\pm\right\}  \,,
\end{equation}
where $N \to \infty$ corresponds to the full single-protein solution. If the boundary conditions on $u$ along the bilayer-protein interfaces are such that $y \to - y$ in Fig.~\ref{figBiPol} we have, by symmetry, that $B_n^\pm=0$ for $n \geq 1$.

Substitution of Eq.~(\ref{constructGenSol}) with Eq.~(\ref{genSolSingle2})
into Eq.~(\ref{genSol}) yields the solution of the thickness deformations
induced by two membrane proteins. To use Eq.~(\ref{EvalEnergyLine}) to evaluate
the elastic energy associated with these thickness deformations, and to impose
suitable boundary conditions along the bilayer-protein interfaces, we recast---along the bilayer-protein boundary associated with protein 2---$\bar u_1(r_1,\theta_1)$ in terms of $r_2$, $\sin \theta_2$, and $\cos \theta_2$, and vice versa. For protein 2, this is achieved \cite{mathematica11} by first expanding the bipolar coordinate transformations for $r_1$, $\sin \theta_1$, and $\cos \theta_1$, and then the expression for $\bar u_1(r_1,\theta_1)$ in Eq.~(\ref{constructGenSol}), in terms of $r_2^\prime = r_2/d$ up to some order $M$ in $r_2^\prime$. Steric constraints mandate $d>R_1+R_2$ and, hence, $r_2^\prime<1$ along the bilayer-protein boundary associated with protein 2. Following a similar procedure for protein 1, Eq.~(\ref{genSol}) yields explicit expressions for $\bar u$ in terms of $(r_2,\theta_2)$ in the vicinity of protein 2 and in terms of $(r_1,\theta_1)$ in the vicinity of protein 1. We note that expansion of $u_1(r_1,\theta_1)$ around protein
2 up to order $M$ in $r_2^\prime$ produces angular variations in $\theta_2$
up to $\sin M \theta_2$ and $\cos M \theta_2$. We set $M=N$ to ensure that these ``secondary''
angular
variations, which are introduced into the general solution in Eq.~(\ref{genSol}) via expansion of the bipolar coordinate transformations, are of the same
maximum order as the angular
variations captured directly by the Fourier-Bessel series in Eq.~(\ref{genSolSingle2}) \cite{footnote}.

The expansions described above yield explicit expressions for $\bar u$ in terms of $(r_1,\theta_1)$ and $(r_2,\theta_2)$ in the vicinity of proteins 1 and 2, respectively. Thus, the expression for the thickness deformation energy in Eq.~(\ref{EvalEnergyLine}) can be written as
\begin{eqnarray}
G&=& -\frac{1}{2} R_1 \int_0^{2 \pi} d \theta_1 g_1(\theta_1)
-\frac{1}{2} R_2 \int_0^{2 \pi} d \theta_2 g_2(\theta_2)\,,\nonumber \\&& \label{intGder}
\end{eqnarray}
where the overall minus signs arise because the bilayer normal vectors point towards decreasing $r_i$ along the bilayer-protein interfaces, and the boundary
energy densities
\begin{eqnarray} \label{Eanaytic1}
g_1(\theta_1)&=& \left[K_b \frac{\partial \bar
u}{\partial r_1} \nabla_1^2 \bar u-K_b \bar u  \frac{\partial}{\partial
r_1}\nabla_1^2 \bar u+\tau \bar u \frac{\partial \bar u}{\partial r_1}\right]_{r_1=R_1}\,,\nonumber\\&&\\
\label{Eanaytic2}
g_2(\theta_2)&=& \left[K_b \frac{\partial \bar
u}{\partial r_2} \nabla_2^2 \bar u-K_b \bar u  \frac{\partial}{\partial
r_2}\nabla_2^2 \bar u+\tau \bar u \frac{\partial \bar u}{\partial r_2} \right]_{r_2=R_2}\nonumber\\&&
\end{eqnarray}
are evaluated using $\bar u(r_1,\theta_1)$ and $\bar u(r_2,\theta_2)$,
respectively, in which we have noted that $\left|\mathbf{\hat n} \cdot \nabla\right|=\frac{\partial}{\partial
r}$ along the circumference of a circle, and the Laplace operators
\begin{equation}
\nabla_i^2 = \frac{\partial^2}{\partial r_i^2}+\frac{1}{r_i} \frac{\partial}{\partial r_i}+\frac{1}{r_i^2} \frac{\partial^2}{\partial \theta_i^2}
\end{equation}
in polar coordinates, where $i=1,2$. As a result of our expansions of the bipolar coordinate
transformations, $\bar u$ around proteins 1 and 2 only depends on $\theta_{1,2}$ through linear sums over $\sin j \theta_{1,2}$ and $\cos j \theta_{1,2}$
with $j \geq 0$, and it is therefore straightforward \cite{mathematica11} to analytically evaluate the angular integrals in Eq.~(\ref{intGder}), resulting in an algebraic expression for $G$.

In the case of two interacting membrane proteins, the general solution in Eq.~(\ref{genSol}) contains the $4 (2N+1)$ coefficients $A_{i,0}^\pm$, $A_{i,n}^\pm$, and $B_{i,n}^\pm$ with $i=1,2$ and $n=1,\dots,N$. These coefficients are determined by the
boundary conditions through the linear system of equations
\begin{equation} \label{matrixEq}
\mathbf{M} \mathbf{c} = \mathbf{b}\,,
\end{equation}
where the vector $\mathbf{c}$ is of length $4 (2N+1)$ and contains each coefficient appearing in Eq.~(\ref{genSol}) as a separate element. The vector $\mathbf{b}$ contains
the boundary conditions on $\bar u$ and its radial derivatives, at proteins 1 and 2, at each order in $\sin j \theta$ and $\cos j \theta$ with $j \geq
0$ as separate elements.
At $j=0$, we have two boundary conditions for each protein, yielding four elements in $\mathbf{b}$. Similarly, for $j>1$ we have four boundary conditions
at each order in $j$ for each protein yielding, in total, the $4 (2N+1)$
independent boundary conditions required to fix the $4 (2N+1)$ independent
coefficients appearing in Eq.~(\ref{genSol}). Finally, the rows of the matrix $\mathbf{M}$ are constructed from the coefficients of $\sin j \theta$ and $\cos j \theta$ with $j \geq 0$ in Eq.~(\ref{genSol}) and their radial derivatives,
at proteins 1 and 2, with each column corresponding to a particular coefficient. For two proteins one therefore obtains four rows at $j=0$ and eight rows at each $j>0$ yielding, as required, a square matrix of order $4 (2N+1)$.

Solution of Eq.~(\ref{matrixEq}) for the coefficients $\mathbf{c}$ \cite{mathematica11}
yields \cite{CAH2013a},
as $N \to \infty$, the exact thickness deformations $u$ in Eqs.~(\ref{transfu})
and~(\ref{genSol}) induced by arbitrary protein configurations and, via Eq.~(\ref{intGder}), the associated thickness deformation energy $G$. However, to make the above solution procedure analytically tractable it is, in practice, necessary to truncate the respective Fourier-Bessel series, and expansions of the bipolar coordinate transformations, at some finite value of $N$. Such
a truncation relies on the assumption that, beyond $N$, angular variations
in Eq.~(\ref{genSol}) can be neglected. The validity of this assumption for a given value of $N$ can be confirmed \cite{CAH2013a} by systematically including higher-order terms. For large $N$, it can be convenient to substitute numerical values for all model parameters, and to numerically solve for $\mathbf{c}$ in Eq.~(\ref{matrixEq}).

The above analytic solution procedure can be implemented directly for the
boundary conditions associated with the cylinder and crown models of membrane proteins discussed in Secs.~\ref{secCylinder} and~\ref{secCrown}. For the clover-leaf model discussed in Sec.~\ref{secClover},
suitable (approximate) boundary conditions can be obtained perturbatively
from Eq.~(\ref{boundCclover})
\cite{CAH2013a,CAH2013b} by expanding the left-hand sides of Eqs.~(\ref{bc1f}) and~(\ref{bc2f}) in terms of the small parameter $\epsilon_i$ \cite{kim00}, and noting that
\begin{equation} \label{trigIdentTrimer}
\cos s \left(\theta_i-\omega_i \right)=\cos s \theta_i \cos s \omega_i+ \sin  s \theta_i \sin s \omega_1\,.
\end{equation}
To leading order in $\epsilon_i$, Eqs.~(\ref{bc1f}) and~(\ref{bc2f}) are
then given~by
\begin{align} \label{genBC1trimerFin}
&\bar u(R_i,\theta_i)+F_i(R_i) \cos s \theta_i+G_i(R_i) \sin s \theta_i=U_i+\frac{\tau
a}{K_t}\,,\\
&\frac{\partial \bar u(r_i,\theta_i)}{\partial r_i}\bigg|_{r_i=R_i}+F_i^\prime(R_i) \cos s \theta_i+G_i^\prime(R_i) \sin s \theta_i=U_i^\prime\,, \label{genBC2trimerFin}
\end{align}
where
\begin{eqnarray} \label{genBC1trimerFin2}
F_i(r_i)&=&R_i \epsilon_i   \frac{\partial \bar u(r_i,\theta_i)}{\partial r_i}  \cos s \omega_i\,, \\ \label{genBC2trimerFin2}
G_i(r_i)&=&R_i \epsilon_i   \frac{\partial \bar u(r_i,\theta_i)}{\partial r_i}  \sin s \omega_i\,.
\end{eqnarray}
Note that, for protein configurations which satisfy $\sin s \omega_i=0$, only cosine
modes must be considered in the above expressions because, for such configurations, the arrangement in Fig.~\ref{figBiPol} is symmetric under $y \to -y$. For the sake of simplicity, we approximate $\bar u$ in Eqs.~(\ref{genBC1trimerFin2}) and~(\ref{genBC2trimerFin2}) by only including the rotationally symmetric ``background'' fields about proteins 1 and 2 in Eq.~(\ref{genSol}). Bilayer thickness-mediated interactions in the clover-leaf model of integral membrane proteins can then be analyzed following the same steps as for the cylinder and crown models, but with the additional approximations inherent in Eqs.~(\ref{genBC1trimerFin}) and~(\ref{genBC2trimerFin}).

\section{Numerical solution}
\label{secNumericalSol}

The FE framework provides a versatile numerical approach for handling protein-induced lipid bilayer deformations in crowded membranes. We have developed a general FE scheme \cite{OWC1,OPWC1} for the numerical study of bilayer-protein interactions which allows reliable and efficient minimization of Eq.~(\ref{energy}) for hundreds of interacting integral membrane proteins with complicated shapes and boundary conditions. To complement our FE approach, we have also developed a FD scheme for the minimization of Eq.~(\ref{energy}).
In this section we provide a detailed description of our FE and FD solution
procedures, which permit minimization of Eq.~(\ref{energy}) for effectively arbitrary protein shapes, separations, and orientations.

While the analytic solution described in Sec.~\ref{secAnalyticSol} allows for infinitely large system sizes, numerical solutions are necessarily restricted to finite solution domains. With the exception of special cases such as provided by periodic systems \cite{OPWC1}, numerical minimization of Eq.~(\ref{energy}) therefore relies on the assumption that the energy cost of thickness deformations decays sufficiently rapidly at large distances from the proteins so that finite size effects can be neglected. This assumption is violated for finite membrane tensions in Eq.~(\ref{energy}), in which case the magnitude of $G_1$ in Eq.~(\ref{energyG1}) increases with bilayer area and, indeed, becomes infinite in the limit of infinitely large lipid bilayers. However, $G_1$ does not contribute to the energy cost of protein-induced lipid bilayer thickness deformations. Thus, direct comparisons between analytic and numerical solutions of the energy cost of protein-induced lipid bilayer thickness deformations in the non-interacting as well as interacting regimes can be made, even for
$\tau>0$, by subtracting the (finite) value of $G_1$ associated with a specific choice for the size of the solution domain
from the numerical solution, and (formally) subtracting the corresponding
(infinite) value of $G_1$ from the analytic solution as in Eq.~(\ref{EvalEnergyLine}).
Furthermore, as we discuss below, comparisons between analytic and numerical
results for the thickness deformation field, rather than the thickness deformation energy, over identical (finite) bilayer areas for the analytic and numerical solutions provide an alternative approach for testing our numerical solution
procedures. This approach does not rely on subtracting $G_1$ in Eq.~(\ref{energy2}).

\subsection{Finite elements}
\label{secFE}

The FE framework \cite{shames1985energy,bathe2006} was developed to permit reliable and computationally efficient numerical solutions of boundary value problems involving large computational domains with complicated boundary shapes and boundary conditions. This makes the FE approach well suited for the calculation of lipid bilayer-mediated protein interactions in crowded membranes with many interacting proteins. Using MscL as a model system, we have shown previously \cite{OWC1,OPWC1} that the FE approach makes it feasible to predict directional bilayer-mediated protein interactions in systems composed of hundreds of integral membrane proteins.

While standard FE methods based on Lagrange interpolation functions \cite{shames1985energy} are sufficient to compute the thickness stretch and tension terms in Eq.~(\ref{energy}), the curvature term, being second order in derivatives, requires $C^1$ continuity \cite{bathe2006} and therefore cannot be handled through Lagrange interpolation functions. The discrete Kirchhoff triangle (DKT) method  
offers an elegant and efficient way to circumvent this limitation \cite{Batoz1980,Bathe1981}.
In particular, to bypass $C^1$ continuity, the DKT approach employs a plate
theory allowing for transverse shear deformations, in which case
$C^0$ continuity is sufficient. The Kirchhoff hypothesis of zero transverse shear is then enforced discretely along the edges
of the triangular elements, thus ensuring the conformity of curvatures at element interfaces.

In our FE framework for calculating bilayer thickness-mediated interactions between integral membrane proteins \cite{OWC1,OPWC1} we adopt a hybrid FE approach, in which we combine
the DKT formulation for the bending terms \cite{Batoz1980} with standard Lagrange interpolation for the thickness stretch and gradient terms \cite{shames1985energy}. To derive the stiffness matrix associated with this hybrid FE approach, we rewrite Eq.~(\ref{energy}) in Cartesian coordinates,
\begin{eqnarray}
G &= \frac{1}{2} \int dxdy \left\{ K_b\left(\frac{\partial^2 u}{\partial x^2} + \frac{\partial^2 u}{\partial y^2}\right)^2
      +\frac{K_t}{a^2} u^2 
      \nonumber \right. \\*&\left.
       + \tau \left[\left(\frac{\partial u}{\partial x}\right)^2+\left(\frac{\partial u}{\partial y}\right)^2\right] 
      + \frac{2\tau}{a} u
      \right\} \, .
\label{FEenergy}
\end{eqnarray}
The variation of Eq.~(\ref{FEenergy}) is given by
\begin{equation}
 \delta G = \int dxdy (\delta  \bold{\epsilon} )^T \mathbf{D} \bold{\epsilon} 
 +\int dxdy \frac{\tau}{a} \delta u \, ,
\end{equation}
with the generalized strain vector
\begin{equation}
 \bold{\epsilon}^T = \left[u \quad \frac{\partial u}{\partial x} \qquad \frac{\partial u}{\partial y} \qquad
                           \frac{\partial^2 u}{\partial x^2} \qquad \frac{\partial^2 u}{\partial y^2}   \right]\,,
\end{equation}
and the constitutive matrix
\begin{equation}
 \mathbf{D} =
 \left[
 \begin{array}{ccccc}
  \frac{K_t}{a^2} & 0 & 0 & 0 & 0 \\
  0 & \tau & 0 & 0 & 0 \\
  0 &  0  & \tau & 0 & 0  \\
  0 & 0 & 0 & K_b & K_b \\
  0 & 0 & 0 & K_b & K_b \\
 \end{array}
\right] \, .
\end{equation}

While the displacements $u_1$, $u_2$, and $u_3$ of the corner nodes of each FE triangle are sufficient to define Lagrange interpolation functions, the DKT approach requires nine degrees of freedom per triangle,
\begin{equation}
 \vec{U}^T = [u_1 \; \; \theta_{x1} \; \; \theta_{y1} \; \; u_2 \; \; \theta_{x2} \; \; \theta_{y2} \; \; u_3 \; \; \theta_{x3} \; \; \theta_{y3}]\,,
\end{equation}
where the partial derivatives $\theta_{xi} = u_{i,y}$ and $\theta_{yi}= -u_{i,x}$ with $i=1,2,$ or 3 correspond to rotations at the corner nodes of each FE triangle. We use the strain-displacement transformation matrix
\begin{equation} \label{eqdefB}
 \mathbf{B} = 
  \left[
 \begin{array}{c}
  \mathbf{G}^T \\ \vspace*{0.1cm}
  \mathbf{G}^T_{,x} \\ \vspace*{0.1cm}
  \mathbf{G}^T_{,y} \\ \vspace*{0.1cm}
  \mathbf{H}^T_{x,x}  \\ \vspace*{0.1cm}
  \mathbf{H}^T_{y,y} 
 \end{array}
\right]
\end{equation}
to construct the strain vector $\bold{\epsilon} = \mathbf{B}\mathbf{U}$,
where the linear triangular shape functions $\mathbf{G}$ are given in Ref.~\cite{shames1985energy} and the DKT shape functions $\mathbf{H}$ are given in Ref.~\cite{Batoz1980}.

Finally, the FE thickness deformation energy is obtained by summing over all finite elements,
\begin{equation}
 G_\text{FE} = \sum_{e\in \text{elements}}  \left(
  \mathbf{U}^e)^T (\mathbf{K}^e \mathbf{U}^e +\mathbf{f}^e
 \right) \, ,
 \label{eq:enFE}
\end{equation}
in which the element stiffness matrix $\mathbf{K}^e$ and ``internal tension'' $\mathbf{f}^e$ are given by
\begin{eqnarray}
  \mathbf{K}^e &=& 2A^e \int  d\xi d\eta \mathbf{B}^T \mathbf{D} \mathbf{B}\, , \nonumber \\ \label{eq:stifftension}
 \mathbf{f}^e &=& 2A^e \int  d\xi d\eta \frac{\tau}{a}\mathbf{G}\, .
\end{eqnarray}
The above integrals are performed over the local coordinates $(\xi,\eta)$ using second-order (three points per element) Gaussian quadrature,
and scaled by the area $A^e$ of the element.

\begin{figure}[t!]
\includegraphics[width=\columnwidth]{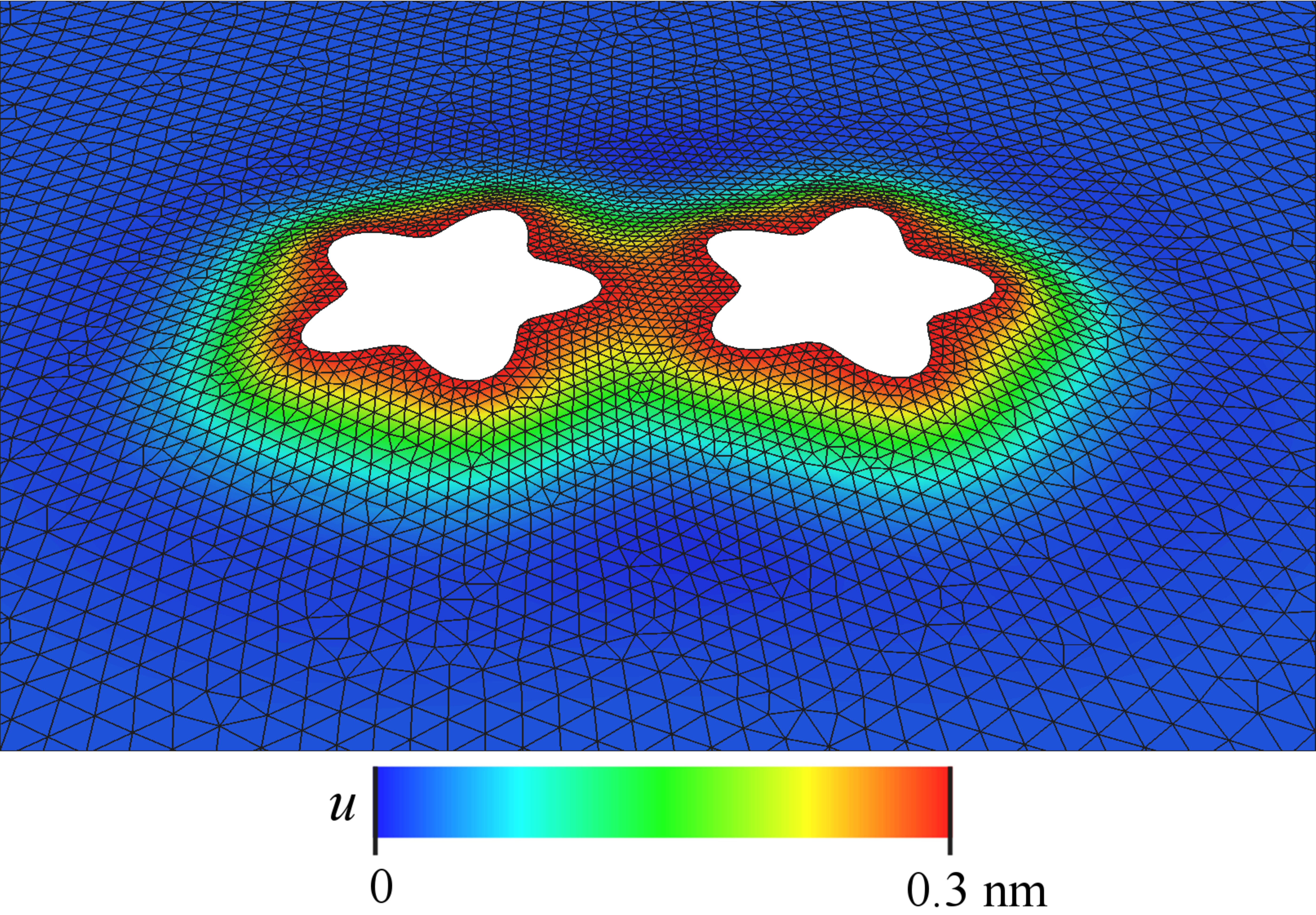}
\caption{(Color online) Thickness deformations $u$ minimizing Eq.~(\ref{energy})
obtained using our FE approach for two clover-leaf pentamers. The triangular mesh (black overlay) indicates the mesh used in the FE calculation, and is generated using the \textsc{frontal} algorithm of the \textsc{gmsh} package \cite{TriangulateRef}.
All model parameters were chosen as described in Sec.~\ref{secElasticModel}.
}
\label{FE1}
\end{figure}

To enforce the general boundary conditions along the bilayer-protein interfaces in Eqs.~(\ref{bc1f}) and~(\ref{bc2f}) we fix $(u, \theta_x, \theta_y)=(U, (U_t)_{,y}, -(U_t)_{,x})$ for the nodes defining the protein boundaries, where $(U_t)_{,x}$ and $(U_t)_{,y}$ are $x$ and $y$ projections of the thickness gradient $U_t$ along the contour of the bilayer-protein interfaces. For the cylinder and clover-leaf models of protein shape the hydrophobic thickness is constant along the bilayer-protein interface, and we therefore
impose $U_t=0$. For the crown model we have $U_t(\theta)=-\delta s \sin(\theta-\omega)$ from the hydrophobic thickness variation in Eq.~(\ref{VarU}). For the nodes defining the outer boundary of the simulation domain we do not constrain $u$ and its derivatives. We assemble the thickness deformation energy in Eq.~(\ref{eq:enFE}) in C++ using the variational mechanics library \textsc{voom} and minimize Eq.~(\ref{eq:enFE}) by employing the \textsc{l-bfgs-b}
solver \cite{Zhu1997}. Figure~\ref{FE1} shows a representative thickness deformation profile obtained for two clover-leaf pentamers together with the corresponding triangular mesh used in the FE calculation, which we generate using the \textsc{frontal} algorithm of the \textsc{gmsh} package \cite{TriangulateRef}.

We check the accuracy of our FE procedure by comparing the total thickness deformation energies predicted by analytic and FE approaches for the cases in which exact analytic results on protein-induced lipid bilayer thickness deformations are available \cite{huang86,CAH2013a,CAH2013b} (see Sec.~\ref{secAnalyticSol}). As discussed above, such comparisons necessitate subtracting $G_1$ in Eq.~(\ref{energyG1}). To complement these tests, we also compare our analytic and FE approaches using the analytic and FE solutions for the thickness deformation field. This approach for quantifying the level of agreement between analytic and FE approaches does not rely on subtracting $G_1$ in Eq.~(\ref{energyG1}). In particular, following Ref.~\cite{Zienkiewicz1987} we monitor the percentage error in the thickness deformations obtained from the FE approximation, $u_h$, relative to the analytic solution, $u$,
\begin{equation} \label{percentageError}
\eta_u = 100 \times \frac{||u-u_h||_{L^2}}{||u||_{L^2}}\,,
\end{equation}
and the corresponding percentage error in curvature deformations,
\begin{equation}
\eta_{\nabla^2 u} = 100 \times \frac{|u-u_h|_{W^{2,2}}}{|u|_{W^{2,2}}} \,,
\label{percentageError2}
\end{equation}
where the $L^2$ norm
\begin{equation} \label{eq:L2}
 ||u||_{L^2} = \left(\int  u^2\; dxdy\right)^\frac{1}{2}
\end{equation}
and the Sobolev semi-norm
\begin{equation} \label{eq:H2}
 |u|_{W^{2,2}} = \left(\int  (\nabla^2u)^2 \;dxdy\right)^\frac{1}{2}
\end{equation}
are evaluated by numerical quadrature 
over identical (finite) bilayer areas for the analytic and numerical solutions,
using an approximately circular integration domain with radius $\approx 22$~nm.

\subsection{Finite differences}
\label{secFD}

\begin{figure}[t!]
\includegraphics[width=0.97\columnwidth]{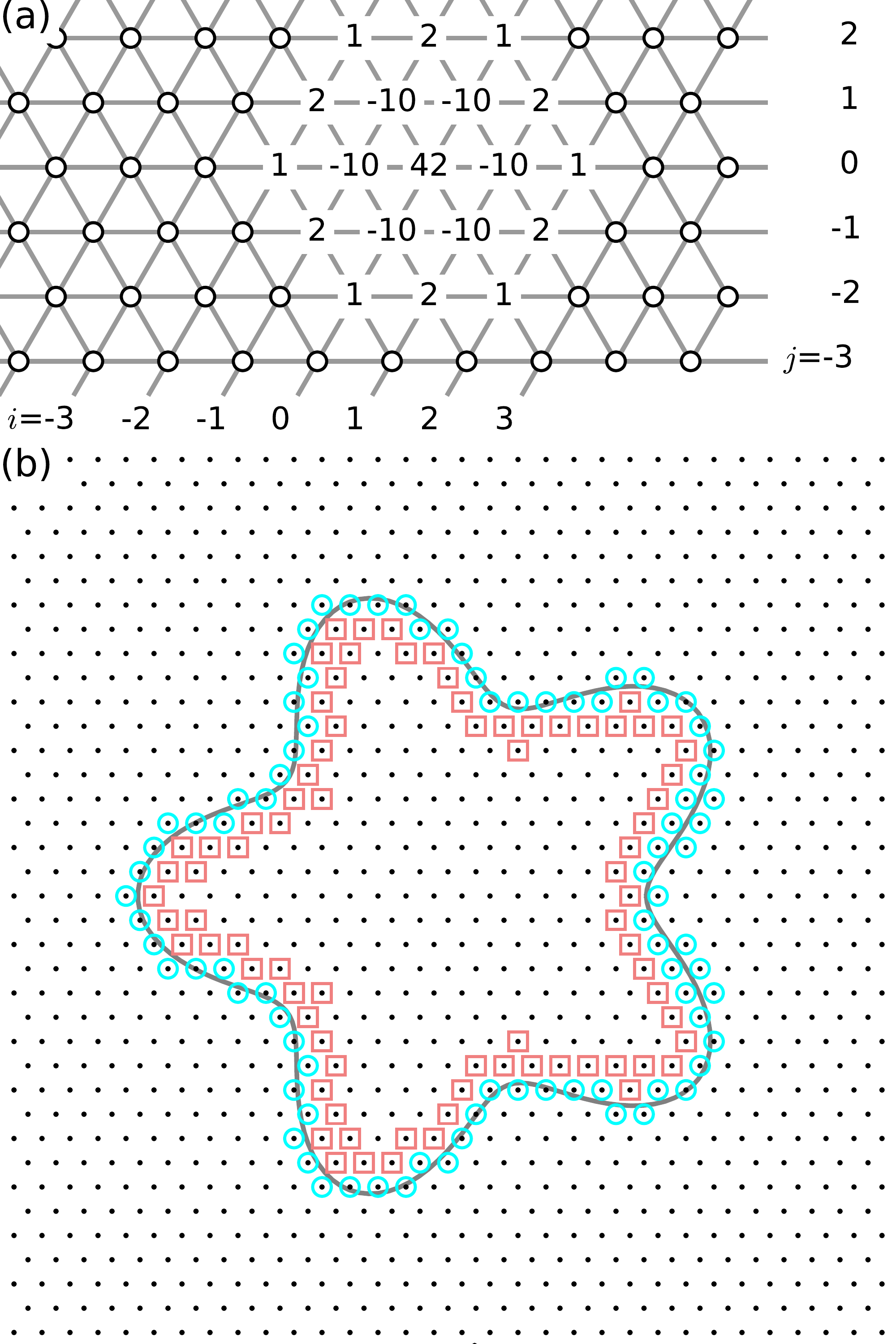}
\caption{(Color online) Illustration of the hexagonal grid used for our FD calculations. (a) Nodal values of the discretized thickness
 deformation field $u_{i,j}$ are indexed by $i$ along the horizontal direction and by $j$
 along the oblique direction. The 19-point stencil discretizing
 the biharmonic operator at position $(i,j)=(0,0)$ is found by multiplying
 $u_{i,j}$ and nearby nodal
 values of the thickness deformation field by the indicated coefficients,
 adding up these contributions, and dividing by $h^4$. (b) We choose the protein boundary points (blue circles) to correspond to the nodes closest to the exact protein boundary curve (grey clover-leaf shape; see Sec.~\ref{secClover}), and impose the general boundary condition in Eq.~(\ref{bc1f}) at these nodes. We use a layer of interior points (red squares), together with corresponding exterior points (see main text), to impose the general boundary condition in Eq.~(\ref{bc2f}).}
 \label{FD1}
\end{figure}

FD methods have been employed previously to study the bilayer thickness deformations induced by gramicidin channels \cite{helfrich90,harroun99b, partenskii03,partenskii04,miloshevsky06}, MscL \cite{ursell07}, G-protein coupled receptors \cite{mondal11,mondal12,mondal14}, and the bacterial leucine transporter \cite{mondal14}. In our FD scheme, we discretize the lipid bilayer
domain of interest using a hexagonal grid with $H\times H$ nodes
and lattice spacing $h$ [see Fig.~\ref{FD1}(a)]. We denote the nodal values of the thickness deformations by $u_{i,j}$. Taylor series expansion then
yields the discretized Laplace operator
\begin{eqnarray}
\textstyle
\nabla^2 u_{i,j} &= &\frac{1}{h^2}\bigg[\frac{2}{3} ( u_{i-1,j+1}+u_{i,j+1}+u_{i-1,j}+u_{i+1,j} \nonumber\\ \label{eqFDLap}
\quad \quad && +u_{i, j-1}+u_{i+1,j-1} -6 u_{i,j}) \bigg]\,.
\end{eqnarray}
Minimization of the thickness deformation energy in Eq.~(\ref{energy}) via solution of the Euler-Lagrange equation~(\ref{genBihpreu}) using FD requires an expression for the discretized biharmonic term $\nabla^4 u_{i,j}$. We obtain $\nabla^4 u_{i,j}$ by applying the discretized Laplace operator in Eq.~(\ref{eqFDLap}) two times, resulting in a 19-point stencil with the coefficients shown in Fig.~\ref{FD1}(a).

For our FD calculations we focus on the case of zero membrane tension, for which Eq.~(\ref{energy}) yields the FD thickness deformation energy
\begin{equation} \label{energyFD}
G_\text{FD} = \frac{\sqrt{3} h^2}{2}\sum_{i,j}\left[ \frac{K_b}{2} \left(\nabla^2 u_{i,j}\right)^2 +\frac{K_t}{2a^2} u_{i,j}^2 \right]\,,
\end{equation}
in which nodal contributions are scaled with the unit area associated with each node. Collecting nodal values of the thickness mismatch $u_{i,j}$ into a vector $\vec{u}$ of length $H^2$, we recast the energy in Eq.~(\ref{energyFD}) into the matrix form
\begin{eqnarray}
 G_\text{FD} &=& \frac{\sqrt{3} h^2}{2}\left[ \frac{K_b}{2}\left(\vec{L} \vec{u}\right)^T (\vec{L}\vec{u}) +\frac{K_t}{2a^2} \vec{u}^T \vec{u} \right] \nonumber \\
  &=& \frac{\sqrt{3} h^2}{2}\left[ \frac{K_b}{2}\left(\vec{u}^T \vec{N} \vec{u}\right) +\frac{K_t}{2a^2} \vec{u}^T \vec{u} \right] \nonumber\\
  &\equiv& \vec{u}^T \vec{Q} \vec{u}\,, \label{CalculateGFD}
\end{eqnarray}
where
\begin{equation}
\vec{Q}=\frac{\sqrt{3} h^2}{2}\left( \frac{K_b}{2} \vec{N} +\frac{K_t}{2a^2} \vec{I} \right)\,,
\end{equation}
and $\vec{L}$, $\vec{N}= \vec{L}^T \vec{L}$, and $\vec{I}$ are $H^2\times H^2$ Laplacian, biharmonic, and identity matrices, respectively. In each row, the matrices $\vec{L}$ and $\vec{N}$ have the coefficients of the discrete Laplace and biharmonic
operators associated with a single node as their elements, respectively. The matrices $\vec{L}$ and $\vec{N}$ are therefore highly sparse, with non-zero elements organized according to the node ordering of the vector $\vec{u}$.

To enforce the general bilayer-protein boundary conditions in Eqs.~(\ref{bc1f}) and~(\ref{bc2f})
in our FD scheme we find all grid points at a distance less than $h/2$ from the protein
boundary, as illustrated in Fig.~\ref{FD1}(b) for clover-leaf shapes. We take these nodes to define the protein boundary in our FD scheme,
and impose the boundary condition in Eq.~(\ref{bc1f}) by setting $u_{i,j}=U$
along the discretized protein boundary. 
To enforce the boundary condition on $\mathbf{\hat n} \cdot \nabla u$ in Eq.~(\ref{bc2f}) we find, for
each protein, the nodes in the interior of the protein boundary curve within a distance $b$ from the protein boundary so that $h/2<b<3h/2$. For each of these interior points we then find a mirror symmetric point exterior to the protein boundary curve such that interior and exterior points are connected by a line of length $2b$ which is normal to the exact protein boundary curve. In order to satisfy the boundary condition $U_i^\prime=0$ used here, we impose
the constraint that the values of $u_{i,j}$ at the interior and exterior points are equal to each other (other
choices for the value of $U_i^\prime$ could be implemented following analogous steps). Here a complication arises in that the exterior points are typically not grid points. We address this issue by interpolating, for each
exterior point, the values of the thickness deformation field at the three nearest grid points.

Due to the substantial computational cost associated with our FD scheme we only employ here the FD approach to study mirror-symmetric protein configurations. For two proteins located at $(\pm d/2,0)$ the size of the solution domain can then be reduced by a factor of one-half by imposing mirror-symmetric boundary conditions along the boundary line $x=0$ (see Fig.~\ref{figBiPol}).
For the remaining boundaries of the solution domain we set $u=0$. For all our FD calculations we use a rectangular solution domain of side lengths ($25$, $25\sqrt{3}/2$)~nm with the protein center placed on the longer midline of the rectangle. We check that our results are robust with respect to increases in the size of the solution domain. To minimize the thickness deformation energy we construct the discretized version of the Euler-Lagrange equation~(\ref{genBihpreu}) for $\tau=0$,
\begin{equation} \label{eqFDEuler}
\vec{Q}  \vec{u}=\vec{v}\,,
\end{equation}
defined for the nodal values $\vec{u}$ of all FD grid points. To enforce the boundary conditions in Eqs.~(\ref{bc1f}) and~(\ref{bc2f}) we adjust the rows of the matrix $\vec{Q}$ in Eq.~(\ref{eqFDEuler}) corresponding to the FD boundary nodes so as to fix the value of $u_{i,j}$ at these nodes, as described above. Accordingly, the vector $\vec{v}$ contains non-zero elements at rows corresponding to the boundary nodes. We solve the linear system in Eq.~(\ref{eqFDEuler}) using the sparse matrix structures and solvers provided by the numerical computing environment \textsc{matlab} \cite{matlab13}. We calculate the corresponding thickness deformation energy using Eq.~(\ref{CalculateGFD}) for the nodal values of all FD grid points lying outside the protein domains.

\section{Cylinder model}
\label{secCylinder2}

In this section we focus on the most straightforward scenario of lipid bilayer
thickness deformations induced by cylindrical membrane proteins [Fig.~\ref{figIllust}(a)], and compare numerical results obtained using our FE and FD schemes to the corresponding exact analytic solutions. As discussed above, we subtract $G_1$ in Eq.~(\ref{energyG1}) to compare thickness deformation energies obtained using numerical
and analytic solution procedures. We first consider the bilayer thickness deformations induced by cylindrical membrane proteins in the non-interacting
regime of large protein separations, and then discuss the bilayer thickness deformations induced by two interacting cylindrical membrane proteins.

\subsection{Non-interacting cylindrical membrane proteins}
\label{secSingleCylind}

\begin{figure}[t!]
\includegraphics[width=\columnwidth]{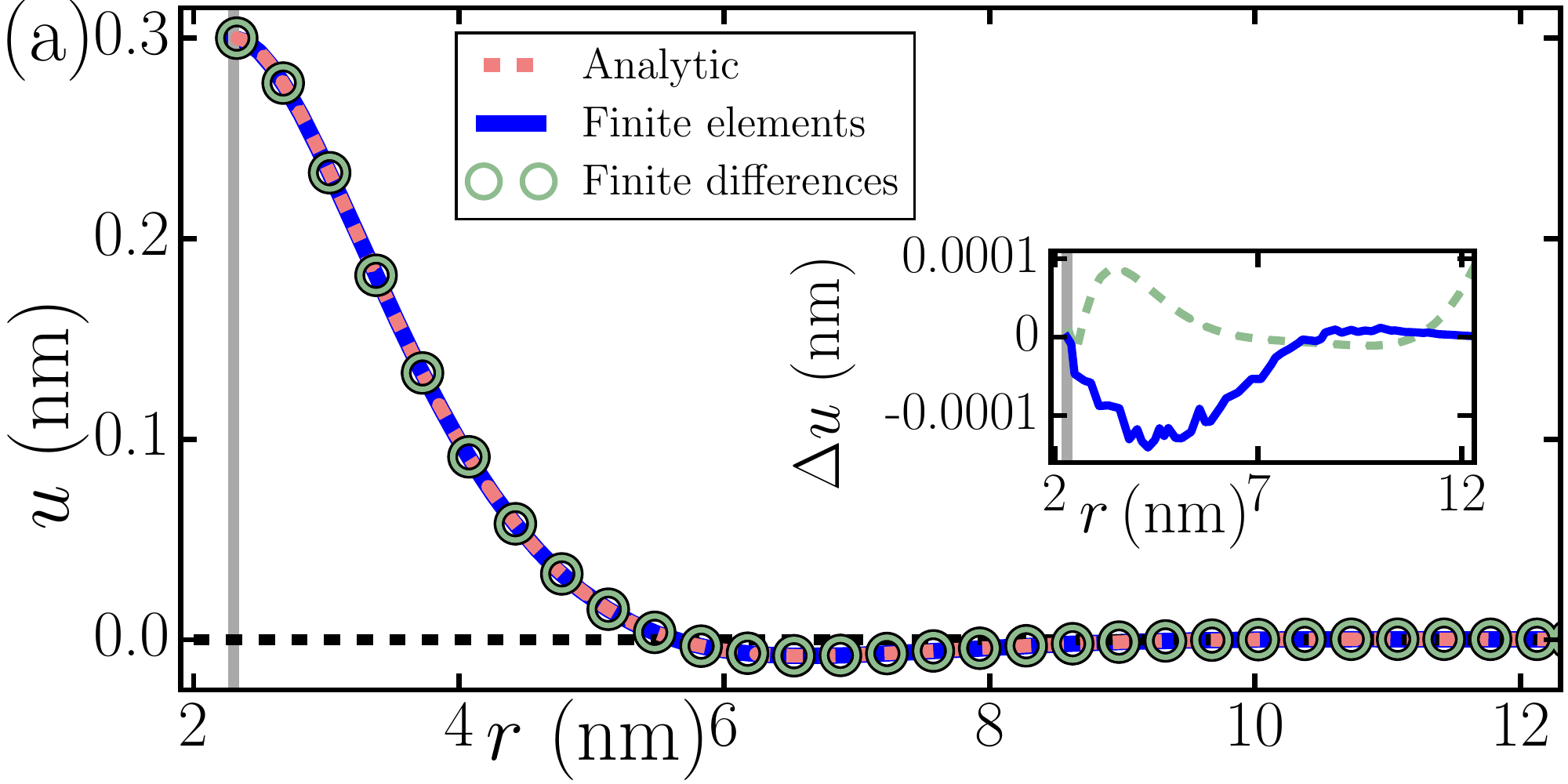}\\ \vspace{0.1cm}
\includegraphics[width=\columnwidth]{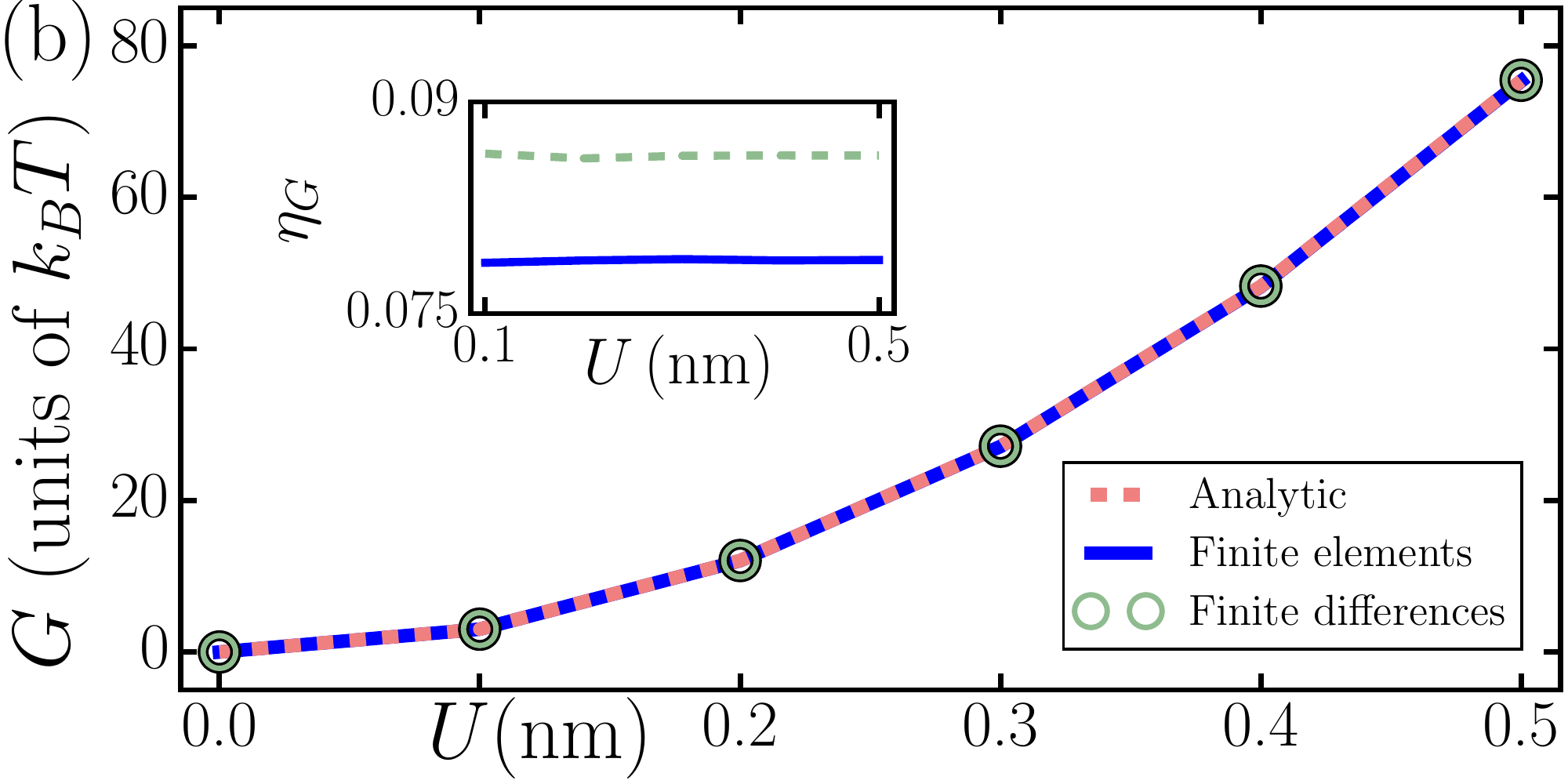}
\caption{(Color online) Lipid bilayer thickness deformations due to a single cylindrical membrane protein. (a) Bilayer thickness deformation profile $u$ versus radial coordinate $r$ obtained from the exact analytic solution in Eq.~(\ref{constructGenSol})
for $d \to \infty$ \cite{huang86}, the FE approach, and the FD approach. The grey vertical line indicates the protein boundary. 
(b) Bilayer thickness deformation energy versus thickness mismatch $U$ obtained using analytic, FE, and FD approaches. 
The insets show (a) the difference in thickness deformation profile, $\Delta u$, and (b) the percentage difference
in thickness deformation energy, $\eta_G$ in Eq.~(\ref{eq:DefetaG}), between the analytic solution and the corresponding results of FE (blue solid curves) and FD (green dashed curves) calculations, respectively.
For the FE solution we used an average edge size of the FE mesh $\langle l_\textrm{edge} \rangle= 0.3$~nm, and for the FD solution we used a lattice spacing $h=0.05$~nm. We set $\tau=0$ for both panels. All model parameters were chosen as described in Sec.~\ref{secElasticModel}, and analytic and numerical solutions were
obtained as discussed in Secs.~\ref{secAnalyticSol} and~\ref{secNumericalSol}.
}
\label{fig:single_cyl} 
\end{figure}

For a single cylindrical membrane protein, the protein-induced lipid bilayer thickness deformations are rotationally symmetric about the protein, with the exact analytic solution \cite{huang86} corresponding to the zeroth-order terms in the general solution in Eq.~(\ref{constructGenSol}). The radial profile
of this exact analytic solution about the membrane protein is governed by zeroth-order modified Bessel functions of the second kind, yielding an approximately exponential decay of thickness deformations with a periodic modulation \cite{huang86,dan93,dan94} and a characteristic length scale of thickness deformations $\lambda=(a^2 K_b/K_t)^{1/4} \approx 1$~nm \cite{phillips09,ursell08} [see Fig.~\ref{fig:single_cyl}(a)]. The thickness deformation profiles obtained using our FE and FD solution procedures are in excellent quantitative agreement with the corresponding analytic solution [Fig.~\ref{fig:single_cyl}(a)]. Computing the percentage difference between numerical and analytic results for the thickness deformation energy,
\begin{equation} \label{eq:DefetaG}
\eta_G = 100\times \left|\frac{G_\text{numerical}-G_\text{analytic}}{G_\text{analytic}}\right|\,,
\end{equation}
we find that the thickness deformation energies obtained using analytic, FE, and FD approaches are also in excellent quantitative agreement [Fig.~\ref{fig:single_cyl}(b)]. As expected from scaling arguments \cite{wiggins04,wiggins05,phillips09},
all three solution procedures yield an approximately quadratic dependence
of the thickness deformation energy on hydrophobic mismatch.

\begin{figure}[t!]
\includegraphics[width=\columnwidth]{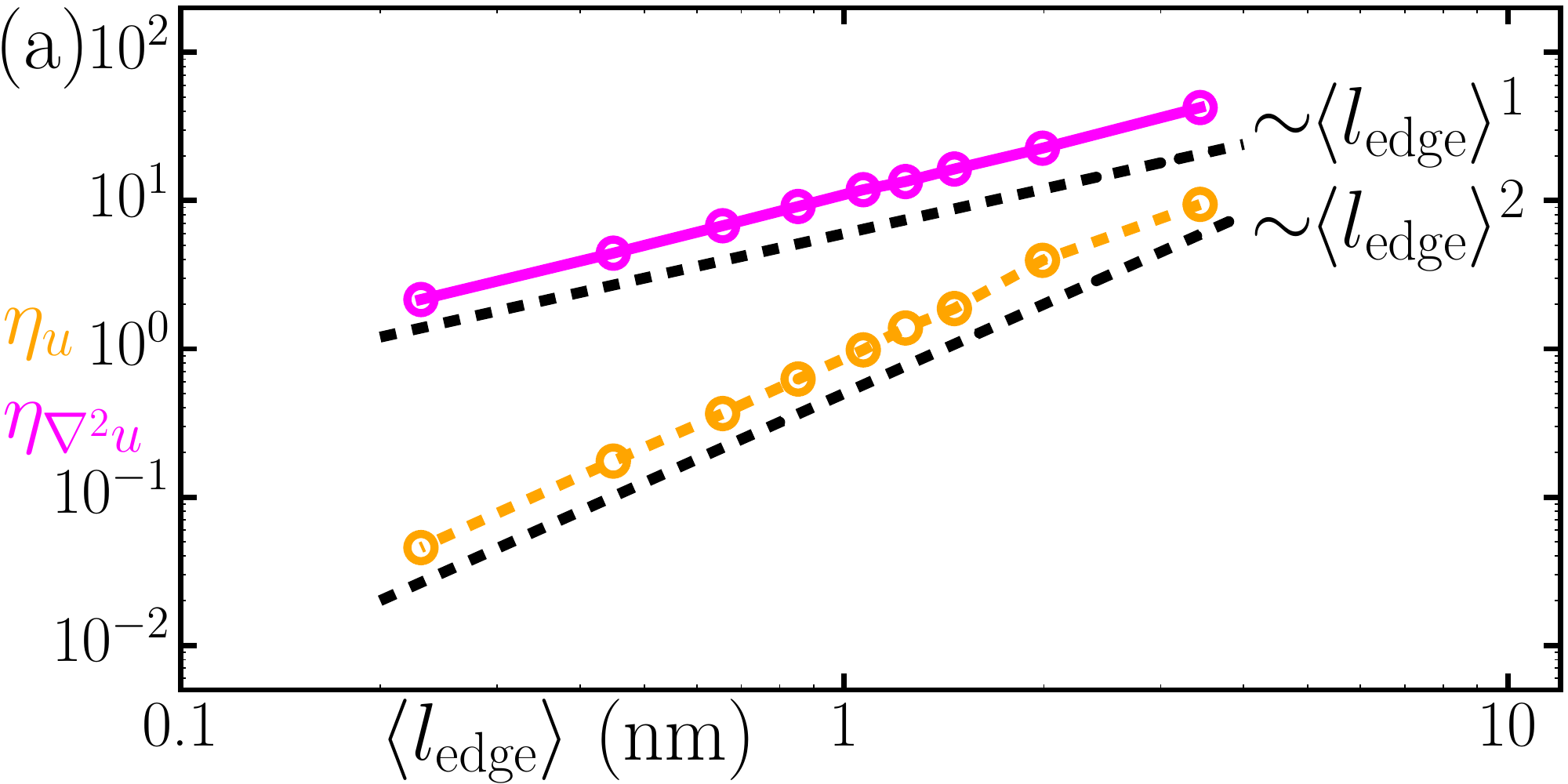}\\ \vspace{0.1cm}
\includegraphics[width=\columnwidth]{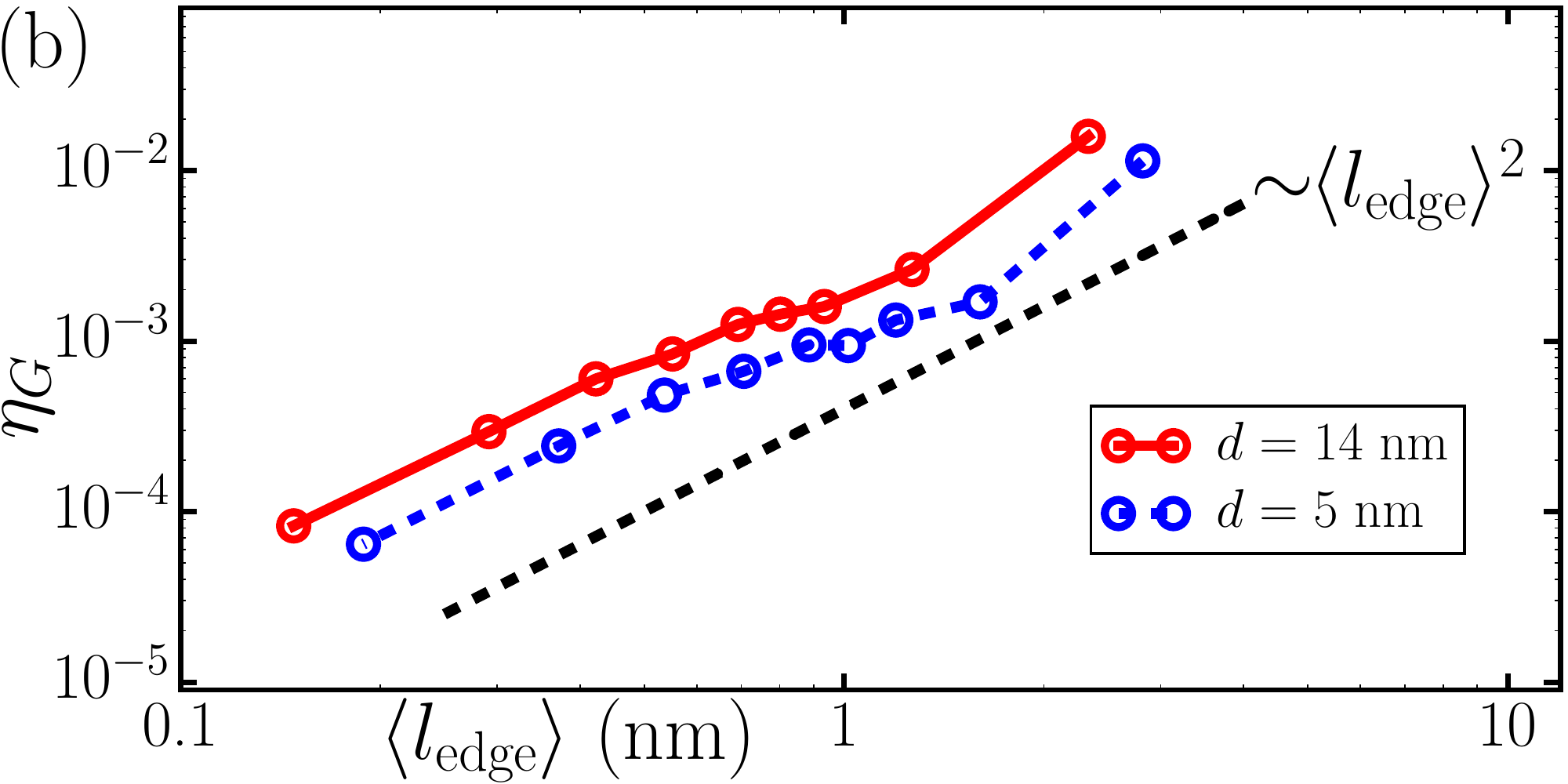}
\caption{(Color online) Comparison of analytic and FE solutions for cylindrical
membrane proteins. (a) Percentage error in thickness deformations in Eq.~(\ref{percentageError})
($\eta_u$; orange dashed curve) and percentage
error in curvature deformations in Eq.~(\ref{percentageError2}) ($\eta_{\nabla^2 u}$; magenta solid curve)  versus $\langle l_\textrm{edge} \rangle$ for a single cylindrical membrane protein. Errors in the thickness mismatch and curvature deformations decay as $\langle l_\textrm{edge} \rangle$ and $\langle l_\textrm{edge} \rangle^2$, respectively. 
(b) Percentage difference between analytic and FE results for the thickness deformation energy per cylindrical membrane protein, $\eta_G$ in Eq.~(\ref{eq:DefetaG}), versus
$\langle l_\textrm{edge} \rangle$ for
a system consisting of two identical cylindrical membrane proteins in the non-interacting regime ($d=14$~nm; red solid curve) and in the strongly interacting regime ($d=5$~nm; blue dashed curve; $N=11$ for the analytic solution). We set $\tau=0$ for both panels. All model parameters were chosen as described in Sec.~\ref{secElasticModel}, and analytic and numerical solutions were
obtained as discussed in Secs.~\ref{secAnalyticSol} and~\ref{secNumericalSol}.
}
\label{fig:cyl_convergence} 
\end{figure}

The convergence of the FE and FD solutions towards the exact analytic solution can be quantified by systematically increasing the spatial resolution of the numerical solution schemes. We find that for the FE solution the percentage errors in thickness and curvature deformations in Eqs.~(\ref{percentageError})
and~(\ref{percentageError2}), which are summed over the entire solution domain, monotonically decrease with decreasing average edge size of the FE mesh, $\langle l_\textrm{edge} \rangle$ [see Fig.~\ref{fig:cyl_convergence}(a)]. In particular, the thickness deformation error decreases approximately quadratically with decreasing $\langle l_\textrm{edge} \rangle$, while the curvature deformation error decreases approximately linearly with decreasing $\langle l_\textrm{edge} \rangle$. As shown in Fig.~\ref{fig:cyl_convergence}(b) (red solid curve), the error in the thickness deformation energy decreases quadratically with decreasing $\langle l_\textrm{edge} \rangle$. While the results in Fig.~\ref{fig:cyl_convergence} were obtained at zero membrane tension, we find that membranes at finite tension yield similar scaling of the errors in the FE thickness deformations, the FE curvature deformations, and the FE thickness deformation energy with $\langle l_\textrm{edge} \rangle$.

\begin{figure}[t!]
\includegraphics[width=\columnwidth]{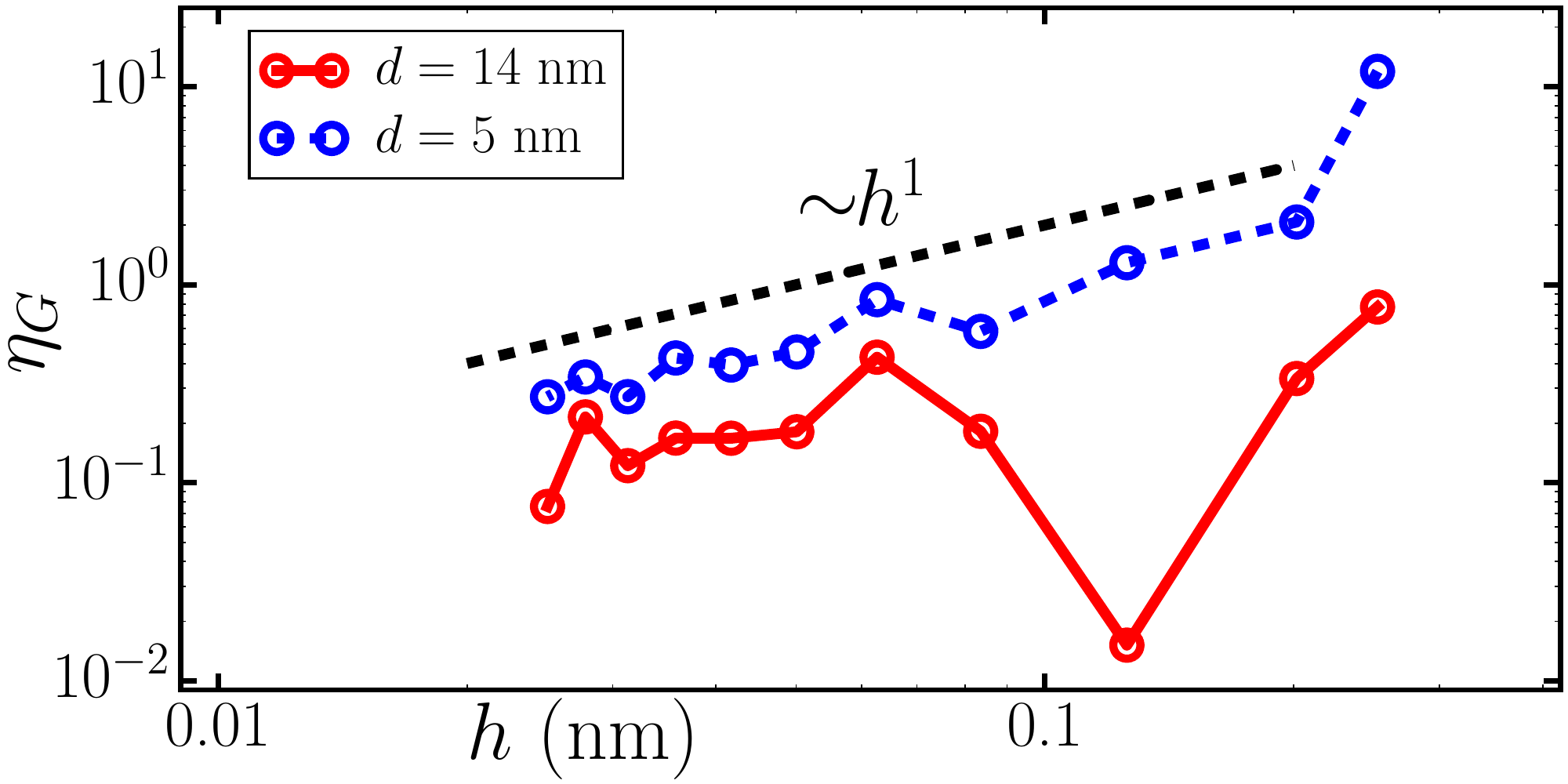}
\caption{(Color online) Percentage difference between analytic and FD results for the thickness deformation energy per cylindrical membrane protein, $\eta_G$
in Eq.~(\ref{eq:DefetaG}), versus lattice spacing, $h$, for a system consisting of two identical cylindrical membrane proteins in the non-interacting regime ($d=14$~nm
for the FD solution; red solid curve) and in the strongly interacting regime ($d=5$~nm; blue dashed curve; $N=11$ for the analytic solution). We set $\tau=0$. All model parameters were chosen as described in Sec.~\ref{secElasticModel}, and analytic and numerical solutions were
obtained as discussed in Secs.~\ref{secAnalyticSol} and~\ref{secNumericalSol}.
} \label{fig:cyl_convergence2} 
\end{figure}

For the FD solution (red solid curve in Fig.~\ref{fig:cyl_convergence2})
the error in the thickness deformation energy decreases approximately linearly with decreasing lattice spacing $h$. The central-difference Laplacian FD stencil we used here is second-order accurate. The linear convergence is
therefore most likely an indication that the FD error is dominated by the enforcement of the slope boundary conditions. Furthermore, we find that, for the parameter values used for Fig.~\ref{fig:cyl_convergence2}, the percentage error in the thickness deformation energy is smaller than $0.5 \%$ for resolutions $h\leq0.2$~nm. Thus, we find that the FE and FD solutions both yield good agreement with the exact analytic solution for high enough spatial resolutions
of the numerical solution schemes. However, Figs.~\ref{fig:cyl_convergence}(b) and~\ref{fig:cyl_convergence2} also show that, compared to the FD solution procedure, the FE solution procedure yields accurate results even at relatively low average spatial resolutions, and converges more rapidly towards the exact analytic result for the thickness deformation energy with increasing average spatial resolution. This suggests that the FE solution procedure is more efficient and, for a given average spatial resolution, more accurate than the FD scheme used here.

\subsection{Interacting cylindrical membrane proteins}

\begin{figure}[t!]
\includegraphics[width=\columnwidth]{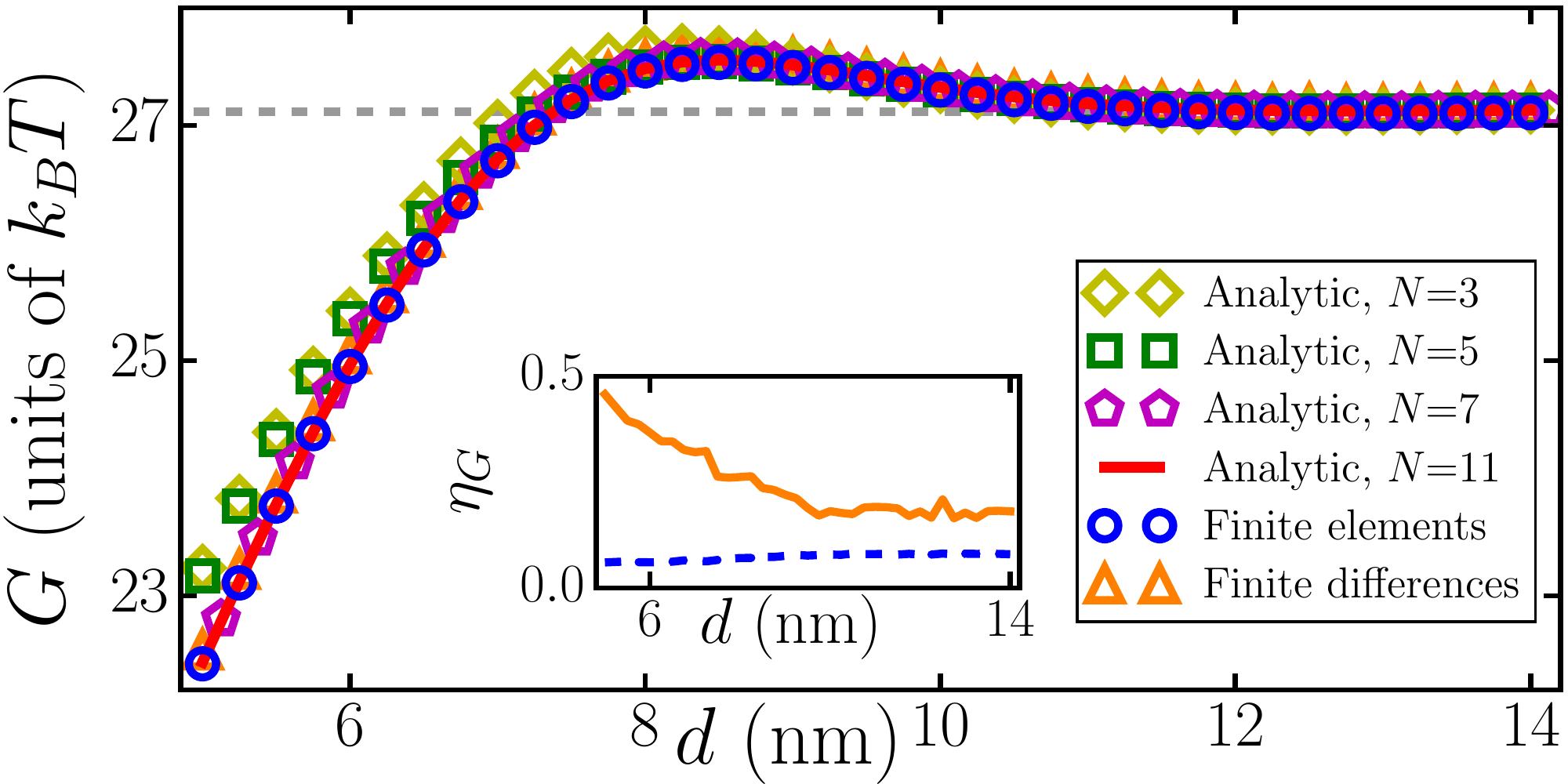}
\caption{(Color online) Thickness deformation energy per protein, $G$, for two cylindrical membrane proteins versus center-to-center protein distance, $d$, calculated analytically at $N=3$, $N=5$, $N=7$, and $N=11$ in Eq.~(\ref{genSol}), and numerically using our FE and FD solution procedures. The inset shows the percentage difference in thickness deformation energy, $\eta_G$
in Eq.~(\ref{eq:DefetaG}), between the analytic solution at $N=11$ and the corresponding results of FE (blue dashed curve) and FD (orange solid curve) calculations. For the FE solution we used $\langle l_\textrm{edge} \rangle \approx 0.27$~nm and for the FD solution we used $h=0.05$~nm. We set $\tau=0$. All model parameters were chosen as described in Sec.~\ref{secElasticModel}, and analytic and numerical solutions were
obtained as discussed in Secs.~\ref{secAnalyticSol} and~\ref{secNumericalSol}.
}
\label{fig:cylinders} 
\end{figure}

For cylindrical membrane proteins of the same hydrophobic thickness, our analytic and numerical solution schemes imply that, for the bilayer and protein parameter values used here, bilayer thickness-mediated interactions are strongly favorable for protein center-to-center distances smaller than $d\approx 8$~nm (see Fig.~\ref{fig:cylinders}). For intermediate protein separations $d\approx8$--$12$~nm, thickness-mediated interactions are weakly unfavorable. We find that thickness-mediated interactions are practically negligible for protein separations greater than $d \approx 12$~nm or minimum protein edge-to-edge separations greater than $\approx 7 \lambda$.

The non-monotonic dependence of bilayer thickness-mediated interactions on protein separation (Fig.~\ref{fig:cylinders}) can be understood \cite{CAH2014a} by considering how the bilayer thickness deformations due to single isolated membrane proteins would interfere. As noted in Sec.~\ref{secSingleCylind}, the thickness deformation profile induced by a single cylindrical membrane protein relaxes rapidly away from the protein, but overshoots with respect to the unperturbed lipid bilayer thickness \cite{huang86,dan93,dan94} [Fig.~\ref{fig:single_cyl}(a)].
Indeed, the zeroth-order modified Bessel functions of the second kind appearing in the general solution in Eq.~(\ref{genSol}) imply successive expansion and compression zones of the lipid bilayer thickness around each protein. When two identical cylindrical membrane proteins are in close proximity to each other, bilayer thickness deformation zones of the same sign overlap and the overall bilayer deformation footprint of the interacting proteins is strongly reduced compared to the large-$d$ limit, resulting in strongly favorable thickness-mediated protein interactions. In contrast, for intermediate protein separations, the protein-induced bilayer thickness deformation profiles are out of phase, yielding substantial overlap of compressed and expanded bilayer regions. This results in frustration of bilayer thickness deformations, and produces weakly unfavorable protein interactions.

As discussed in Sec.~\ref{secAnalyticSol}, the analytic solution of bilayer
thickness-mediated protein interactions in Eq.~(\ref{genSol}) is, in practice, only obtained up to some finite order $n=N$. While modes with $n\geq1$ are irrelevant at large protein separations, modes with non-zero $n$ are essential to correctly account
for the thickness deformations induced by interacting proteins. To confirm the validity of the truncated expansion in Eq.~(\ref{genSol}), we compare the analytic solution of thickness-mediated protein interactions to high-resolution numerical solutions obtained by FE and FD methods (Fig.~\ref{fig:cylinders}). At low orders of the analytic solution, $N=3$ and $N=5$, our analytic estimates of the thickness-mediated interaction energy reproduce the qualitative features of the numerical interaction potentials, but exceed the numerical results by several $k_B T$. Increasing the order of the analytic solution to $N=7$ and $N=11$, we obtain convergence of the analytic solution even for small values of $d$. We find that these high-order analytic solutions are in good quantitative agreement with the corresponding FE solution.

The agreement between FE and high-order analytic solutions is approximately independent of protein separation [Fig.~\ref{fig:cylinders}(inset)]. In particular,
FE and analytic approaches agree similarly well in the non-interacting regime, for which only zeroth-order modes of the analytic
solution must be considered, and in the strongly interacting regime for which, in principle, and infinite number of higher-order modes should be considered in the analytic solution. This suggests that the FE and high-order analytic solutions correctly account for thickness-mediated protein interactions even at very small $d$. Furthermore, we find that, in the strongly interacting regime, the discrepancy between FE and high-order analytic results for the thickness deformation energy decreases approximately quadratically with decreasing $\langle l_\textrm{edge} \rangle$ [blue dashed curve in Fig.~\ref{fig:cyl_convergence}(b)], in agreement with the corresponding result obtained for non-interacting membrane proteins [red solid curve in Fig.~\ref{fig:cyl_convergence}(b)]. While the results in Figs.~\ref{fig:cyl_convergence}(b) and~\ref{fig:cylinders} were obtained with $\tau=0$, we find similar agreement between FE and high-order analytic solutions for finite membrane tensions.

In contrast to the FE solution procedure, the discrepancy between FD and high-order analytic solutions tends to increase with decreasing $d$ [Fig.~\ref{fig:cylinders}(inset)]. This can be understood by noting that, in the FD scheme, very small lattice spacings are required to resolve protein-induced lipid bilayer deformations in the strongly interacting regime.  As in the case of non-interacting proteins (red solid curve in Fig.~\ref{fig:cyl_convergence2}), the discrepancy between FD and high-order analytic results for the thickness deformation energy decreases approximately linearly with decreasing lattice spacing (blue dashed curve in Fig.~\ref{fig:cyl_convergence2}). As in Sec.~\ref{secSingleCylind} this
points to approximate enforcement of slope boundary conditions as the likely dominant source of error in the FD solutions.  Moreover, Fig.~\ref{fig:cyl_convergence2} shows that, consistent with Fig.~\ref{fig:cylinders}, the FD scheme produces larger discrepancies with FE and high-order analytic solutions at smaller $d$ than larger $d$, independent of the lattice spacing considered. For instance, for small $d$ a too coarse lattice spacing $h=0.2$~nm can yield errors in the thickness deformation energy $>2$~$k_B T$, while the same lattice spacing only produces errors $<0.4$~$k_B T$ at large protein separations. In contrast, the convergence of FE and high-order analytic solutions with decreasing $\langle l_\textrm{edge} \rangle$ is not diminished in the interacting regime compared to the non-interacting regime [Fig.~\ref{fig:cyl_convergence}(b)].

\section{Crown model}
\label{secCrownResults}

The cylinder model of integral membrane proteins 
\cite{huang86,wiggins05,ursell08,helfrich90,andersen07,phillips09,nielsen98,nielsen00}
provides a beautiful ``zeroth-order'' description of thickness-mediated protein interactions, but is not able to capture the discrete symmetries and distinct hydrophobic shapes of membrane proteins suggested by membrane structural biology. As discussed in Sec. \ref{secCrown},
a straightforward way to account for rotational asymmetry of the hydrophobic surface of membrane proteins is to allow for angular variations in protein
hydrophobic thickness while maintaining a circular protein cross section \cite{CAH2013a}, resulting in the crown model of membrane proteins [Fig.~\ref{figIllust}(b)]. Angular variations in protein
hydrophobic thickness yield rotationally asymmetric distributions of compression and expansion zones about the protein. For two or more such proteins in close enough proximity, the anisotropy of overlapping deformation fields produces directionality of thickness-mediated protein interactions \cite{CAH2013a}, which is expected to affect protein organization and function.

\begin{figure}[t!]
\includegraphics[width=\columnwidth]{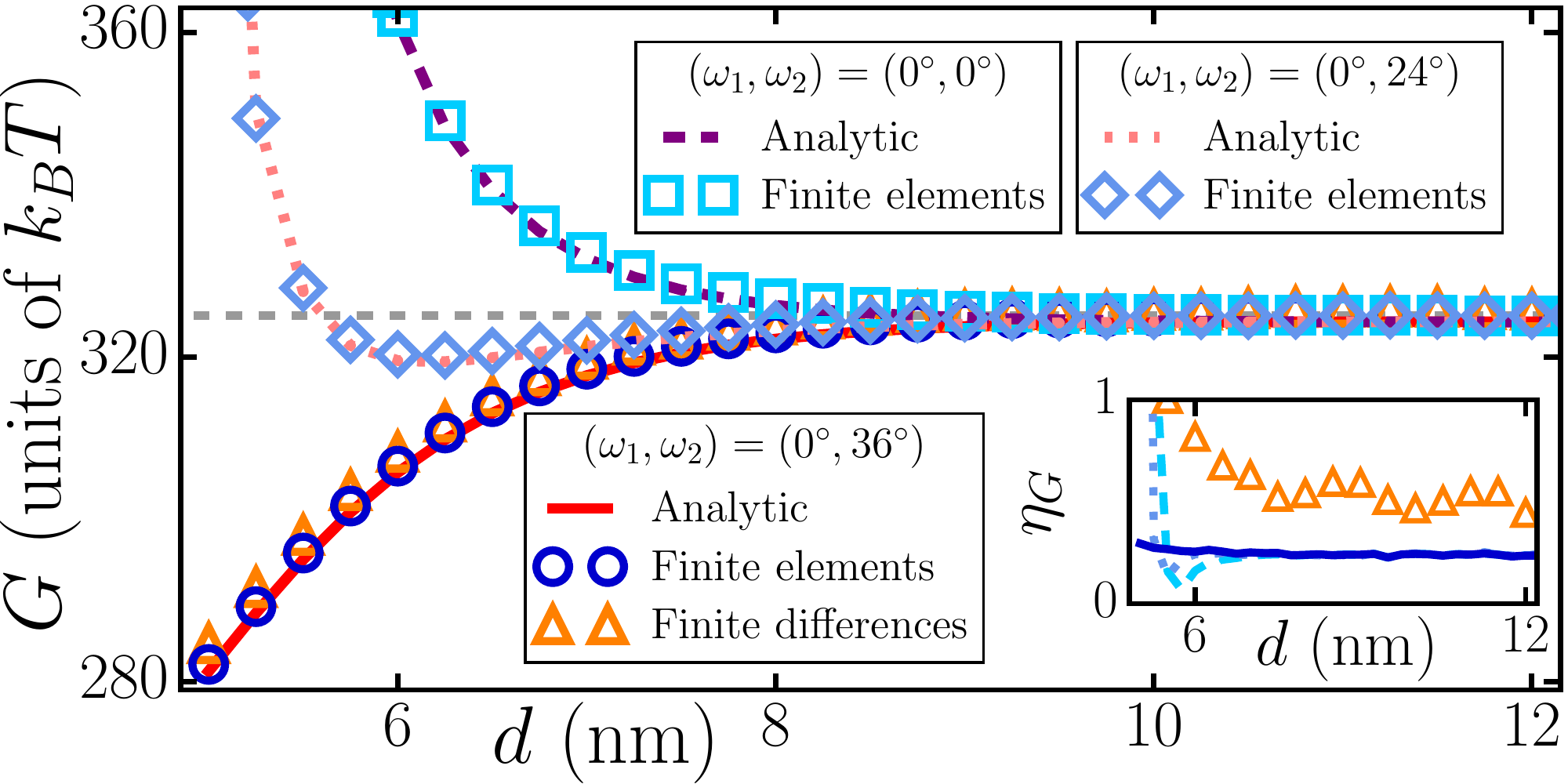}
\caption{(Color online) Thickness deformation energy per protein for two
crown shapes, $G$, versus center-to-center protein distance, $d$, calculated analytically at $N=11$, and numerically using FE
and FD schemes, for the minus-minus configuration [$\omega_1=0^\circ$ and $\omega_2=36^\circ$ in Eq.~(\ref{VarU})], the plus-minus configuration [$\omega_1=0^\circ$
and $\omega_2=0^\circ$ in Eq.~(\ref{VarU})], and a configuration intermediate between the minus-minus and plus-minus configurations [$\omega_1=0^\circ$
and $\omega_2=24^\circ$ in Eq.~(\ref{VarU})]. We only consider here FD solutions for the mirror-symmetric minus-minus protein configuration. The inset shows the percentage difference in thickness deformation energy, $\eta_G$
in Eq.~(\ref{eq:DefetaG}), between the analytic result at $N=11$ and the corresponding results of the FE [blue curves; minus-minus (solid), intermediate (short dashed), and plus-minus (long dashed) configurations] and FD (orange triangles with $\eta_G\approx1.2\%$ at $d=5$~nm) calculations. We used
$\langle l_\textrm{edge} \rangle \approx 0.27$~nm for the FE and $h=0.05$~nm
for the FD solutions, and set $\tau=0$. All model parameters were chosen as described in Sec.~\ref{secElasticModel}, and analytic and numerical solutions were obtained as discussed in Secs.~\ref{secAnalyticSol} and~\ref{secNumericalSol}.
}
\label{fig:crowns} 
\end{figure}

We used our analytic and numerical solution procedures to determine the thickness deformation energy associated with two crown shapes in the ``minus-minus configuration'' [$\omega_1=0^\circ$ and $\omega_2=36^\circ$ in Eq.~(\ref{VarU})], in which protein boundary regions with minimal hydrophobic thickness face each other at the point of closest protein edge-to-edge separation, in the ``plus-minus configuration'' [$\omega_1=0^\circ$ and $\omega_2=0^\circ$ in Eq.~(\ref{VarU}); see Fig.~\ref{figIllust}(b)], in which protein boundary regions with maximal and minimal hydrophobic thickness face each other at the point of closest protein edge-to-edge separation, and in a protein configuration corresponding to $\omega_1=0^\circ$ and $\omega_2=24^\circ$ in Eq.~(\ref{VarU}), which is intermediate between minus-minus and plus-minus configurations (see Fig.~\ref{fig:crowns}).
We find that, in the non-interacting regime, FE and exact analytic solutions are in good quantitative
agreement, with the discrepancy in thickness deformation energy $<0.3\%$ for the parameter values used in Fig.~\ref{fig:crowns} [see Fig.~\ref{fig:crowns}(inset)]. In contrast, our FD solution yields more substantial discrepancies with the exact analytic solution. We attribute this to the difficulty of accurately and unambiguously imposing complicated boundary conditions in the FD scheme.

In the interacting regime, we find that analytic, FE, and FD solutions
predict the same basic qualitative properties of bilayer thickness-mediated interactions between crown shapes (Fig.~\ref{fig:crowns}). Depending on relative protein orientation, thickness-mediated interactions can switch from being strongly favorable to being strongly unfavorable at small protein separations. In particular, when the periodic undulations of the thickness deformations induced
by the two proteins are in phase in the membrane region separating the two proteins, as in the case of the minus-minus configuration in Fig.~\ref{fig:crowns}, similar patterns of protein-induced compression and expansion zones of the lipid bilayer overlap for small $d$, yielding strongly favorable interactions. In contrast, when the thickness undulations induced by the two proteins are out of phase, as in the case of the plus-minus configuration in Fig.~\ref{fig:crowns}, there is substantial overlap of out-of-phase compression and expansion
zones for small $d$, resulting in frustration of bilayer thickness deformations and strongly unfavorable bilayer thickness-mediated interactions. As the relative  protein orientation is changed continuously from in-phase to out-of-phase configurations, the interaction potentials change smoothly \cite{CAH2013a}
from being favorable to being unfavorable at small $d$ which, as in the case of the intermediate configuration in Fig.~\ref{fig:crowns}, can produce a minimum in the thickness-mediated interaction energy at a characteristic protein separation.

On a quantitative level, we find that our FE results on bilayer thickness-mediated interactions between crown shapes are generally in good quantitative agreement
with our analytic solution (Fig.~\ref{fig:crowns}). In contrast, the FD solution procedure yields more substantial discrepancies with our analytic results. In particular, analytic and FE solutions are in good quantitative agreement for all configurations in Fig.~\ref{fig:crowns} with $d>5.3$~nm, with a discrepancy
in thickness deformation energy $<0.5\%$. We find a similar level of quantitative
agreement between analytic and FE solutions even for the smallest protein separations considered for the minus-minus configuration in Fig.~\ref{fig:crowns}.

\begin{figure}[t!]
\includegraphics[width=\columnwidth]{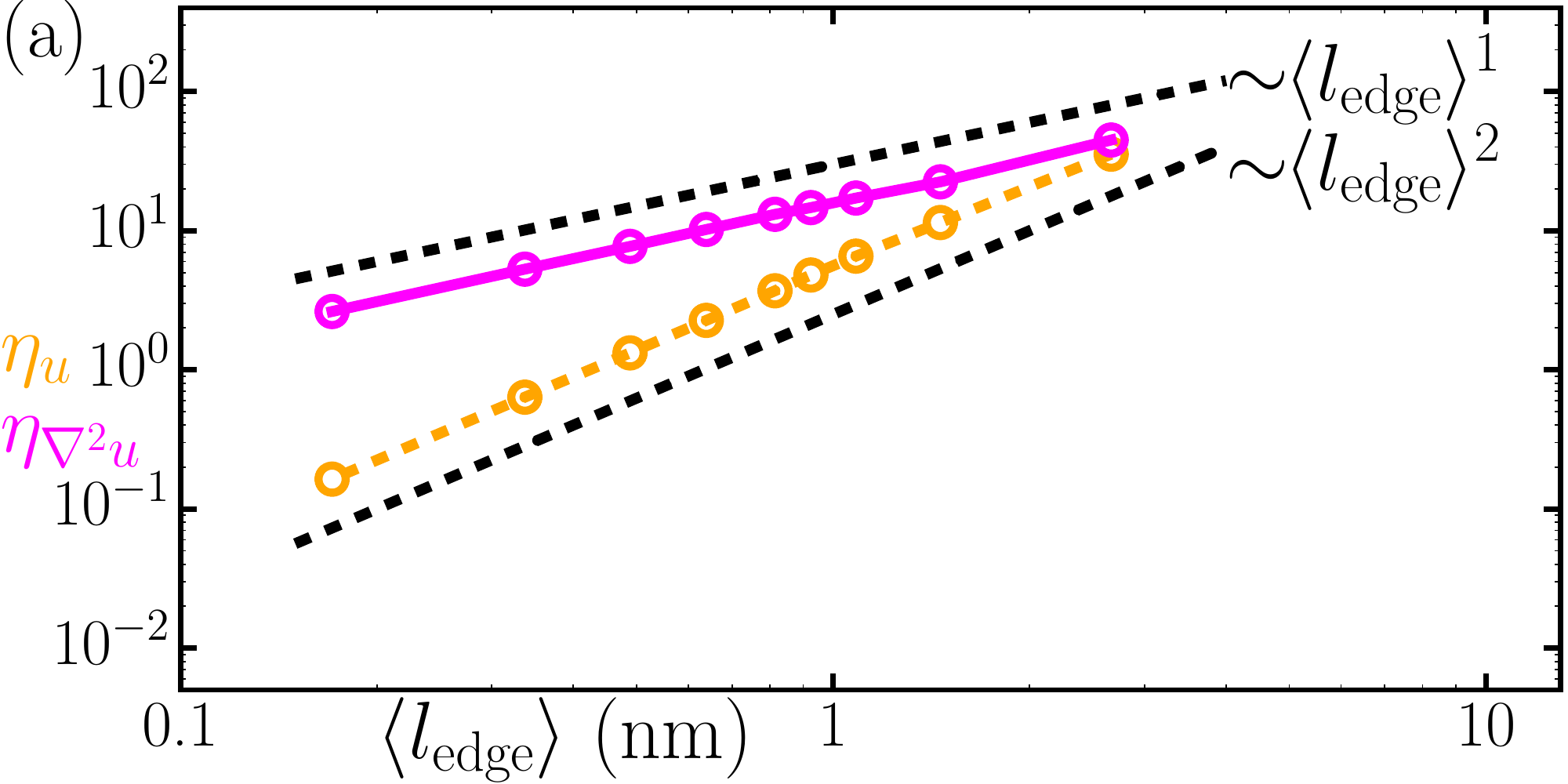}\\
\includegraphics[width=\columnwidth]{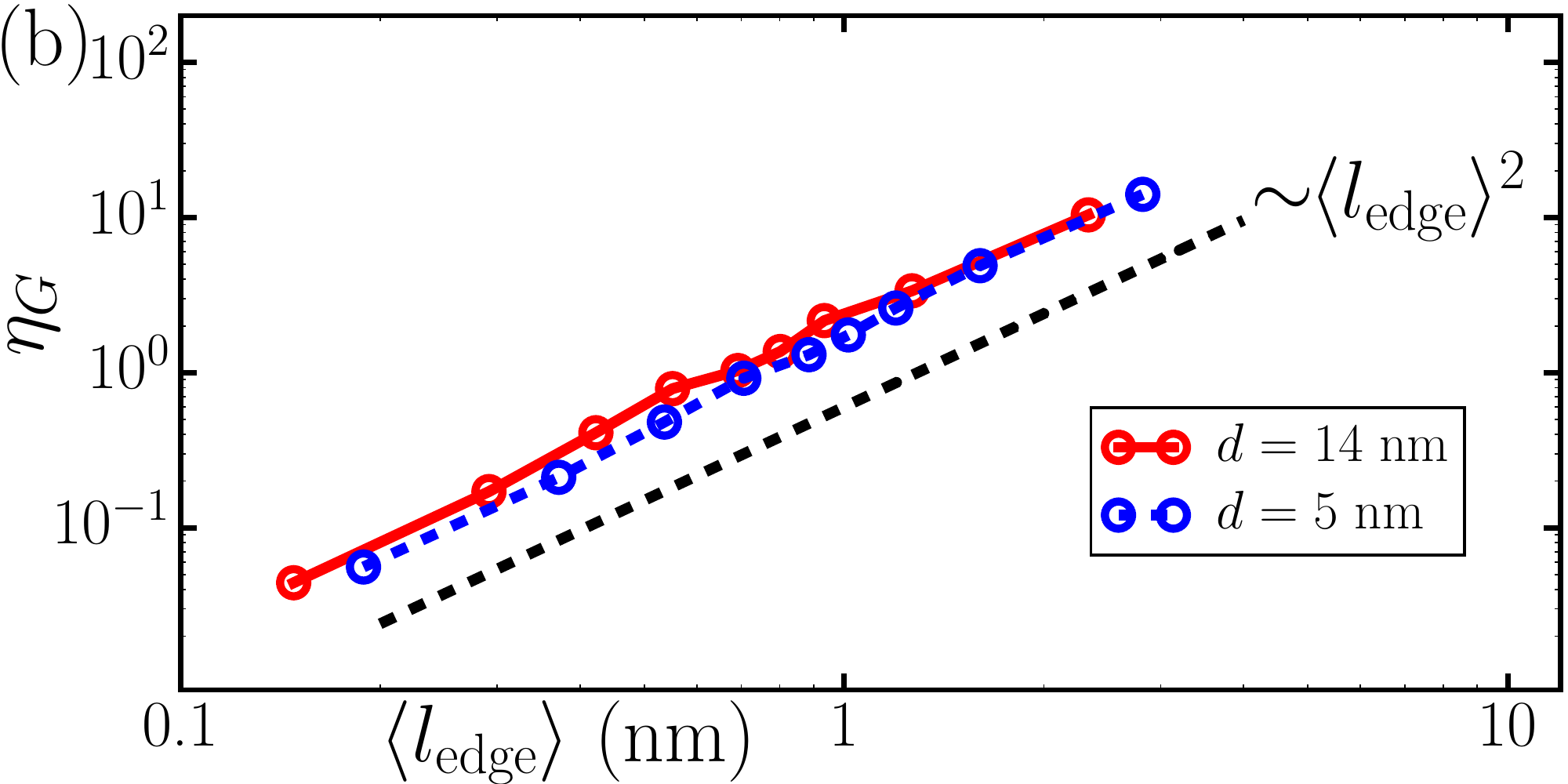}
\caption{(Color online) Comparison of analytic and FE solutions for crown
shapes. (a) Percentage error in thickness deformations in Eq.~(\ref{percentageError})
($\eta_u$; orange dashed curve) and percentage error in curvature deformations in Eq.~(\ref{percentageError2}) ($\eta_{\nabla^2 u}$; magenta solid curve) versus $\langle l_\textrm{edge} \rangle$ for two crown shapes in the minus-minus configuration
in the interacting regime at $d=7$~nm. The error is evaluated using the corresponding analytic result
at $N=11$. Errors in the thickness mismatch and curvature deformations decay as $\langle l_\textrm{edge} \rangle$ and $\langle l_\textrm{edge} \rangle^2$, respectively. 
(b) Percentage difference between analytic and FE results for the thickness deformation energy per protein, $\eta_G$ in Eq.~(\ref{eq:DefetaG}), versus $\langle l_\textrm{edge} \rangle$ for two crown shapes in the minus-minus configuration in the non-interacting regime ($d=14$~nm for the FE solution; red solid curve) and in the strongly interacting regime ($d=5$~nm; blue dashed curve; $N=11$ for the analytic
solution). We set $\tau=1$~$k_B T/\textrm{nm}^2$. All model parameters were chosen as described in Sec.~\ref{secElasticModel}, and analytic and numerical solutions were obtained as discussed in Secs.~\ref{secAnalyticSol} and~\ref{secNumericalSol}.
\label{fig:crown_convergence}}
\end{figure}

The discrepancy between analytic and FE results is most pronounced at very small $d$ in the strongly unfavorable regime of bilayer thickness-mediated interactions, $d<5.3$~nm, for the plus-minus and intermediate configurations in Fig.~\ref{fig:crowns}, where the interaction energy diverges. However, even in this regime the percentage difference between analytic and FE results is $<18\%$ for the plus-minus and intermediate configurations in Fig.~\ref{fig:crowns}, for which the thickness deformation energy can exceed $G\approx 1000$~$k_B T$ for the smallest value $d \approx 5$~nm we allow here. We find that in the strongly unfavorable regime $d<5.3$~nm the magnitude of the gradient of $u$ can exceed $\| \nabla u \|=3$. The maximum value of $\| \nabla u \|$ in the strongly interacting regime is therefore substantially greater than the maximum magnitude of the gradient of bilayer thickness deformations $\approx1$ in the non-interacting regime of the crown shapes considered here, which
we take to induce large gradients of the thickness deformation field so as to test the mathematical limits of applicability of our analytic and numerical solution procedures (see Sec.~\ref{secCrown}). The more pronounced discrepancy between FE and analytic results in the strongly unfavorable regime of bilayer thickness-mediated interactions between crown
shapes arises
because, in this regime, bilayer thickness deformations show a strong variation over the small membrane region separating the two proteins. High accuracy in the
strongly unfavorable regime of thickness-mediated interactions therefore
requires highly refined meshes in the FE scheme and, to capture pronounced angular variations in the thickness deformation field, a large value of $N$
in the analytic approach.

\begin{figure}[t!]
\includegraphics[width=\columnwidth]{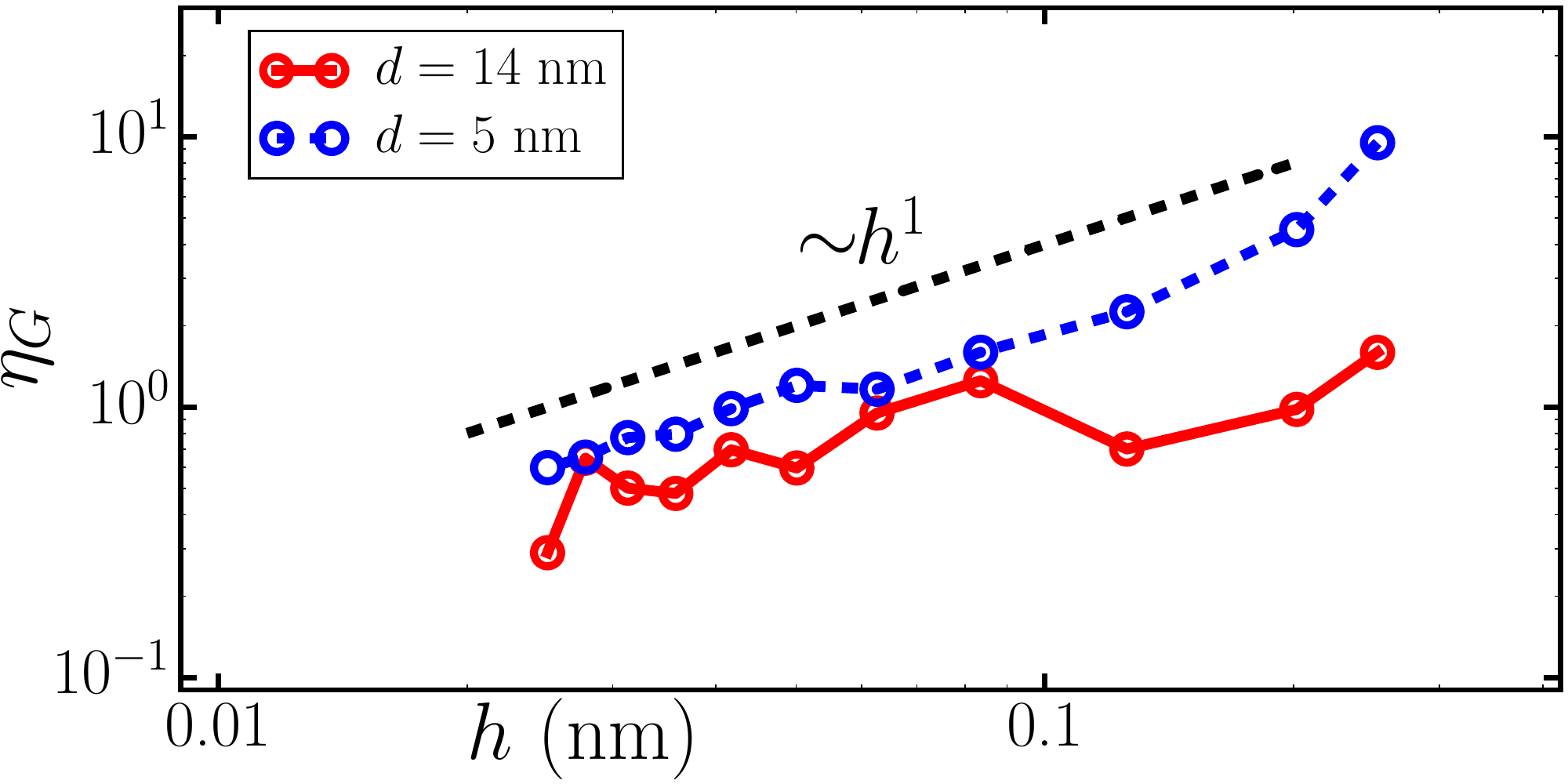}
\caption{(Color online) Percentage difference between analytic and FD results for the thickness deformation energy per protein, $\eta_G$ in Eq.~(\ref{eq:DefetaG}), versus lattice spacing, $h$, for two crown shapes in the plus-plus configuration in the non-interacting regime ($d=14$~nm for the FD solution; red solid curve) and in the strongly interacting regime ($d=5$~nm; blue dashed curve; $N=11$ for the analytic
solution). We set $\tau=0$. All model parameters were chosen as described in Sec.~\ref{secElasticModel}, and analytic and numerical solutions were
obtained as discussed in Secs.~\ref{secAnalyticSol}~and~\ref{secNumericalSol}.
}
\label{fig:crown_convergence2} 
\end{figure}

Following Sec.~\ref{secCylinder2} we quantify the discrepancy between numerical and analytic solutions by systematically increasing the spatial resolution of the numerical solutions. As in the case of cylindrical membrane proteins we find that, for the FE solution procedure, the errors in thickness deformations and thickness
deformation energy decrease approximately quadratically with decreasing average edge size of the FE mesh, while the error in curvature deformations decreases approximately linearly with $\langle l_\textrm{edge} \rangle$ (see Fig.~\ref{fig:crown_convergence}).
We find similar scaling of the discrepancy between FE and analytic solutions with $\langle l_\textrm{edge} \rangle$ in the interacting and non-interacting regimes, as well as for zero and finite membrane tensions.

As for cylindrical membrane proteins, the error in the thickness deformation energy obtained from the FD solution procedure decreases approximately linearly with decreasing lattice spacing in the interacting and non-interacting regimes (see Fig.~\ref{fig:crown_convergence2}), which again points to approximate enforcement of slope boundary conditions as the likely dominant source of error in the FD solutions. Furthermore, Fig.~\ref{fig:crown_convergence2} suggests that, similar to the case of cylindrical membrane proteins, the FD scheme generally produces larger discrepancies with high-order analytic solutions at smaller $d$ than larger $d$. In contrast, the convergence of FE and high-order analytic solutions
is not diminished in the interacting regime compared to the non-interacting regime [Fig.~\ref{fig:crown_convergence}(b)]. Finally, comparison of Figs.~\ref{fig:cyl_convergence}
and~\ref{fig:crown_convergence}, and Figs.~\ref{fig:cyl_convergence2} and~\ref{fig:crown_convergence2}, shows that, for a given spatial resolution, the discrepancy between FE and FD solutions and the analytic solution is more pronounced for crown shapes than for cylindrical membrane proteins.

\section{Clover-leaf model}
\label{secCloverResults}

The clover-leaf model of integral membrane proteins discussed in Sec.~\ref{secClover} can be used to capture non-circular bilayer-protein boundary curves [Fig.~\ref{figIllust}(c)], and provides a generalization of the cylinder model of membrane proteins
complementary to the crown model. In particular, to model the discrete symmetries of integral membrane proteins suggested by membrane structural biology, the clover-leaf model allows for periodic modulations in the shape of the hydrophobic cross section of membrane proteins. As a result, the bilayer thickness deformations induced by clover-leaf proteins show a characteristic pattern of compression and expansion zones about the protein \cite{CAH2013b}, which
provides a simple description of the effect of protein shape and oligomeric state on lipid bilayer thickness deformations. Furthermore, for two clover-leaf proteins in close enough proximity, bilayer thickness-mediated interactions are directional \cite{CAH2013a,OWC1,CAH2014a,OPWC1} and bear characteristic signatures of protein shape, symmetry, and orientation.

\begin{figure}[t!]
\includegraphics[width=\columnwidth]{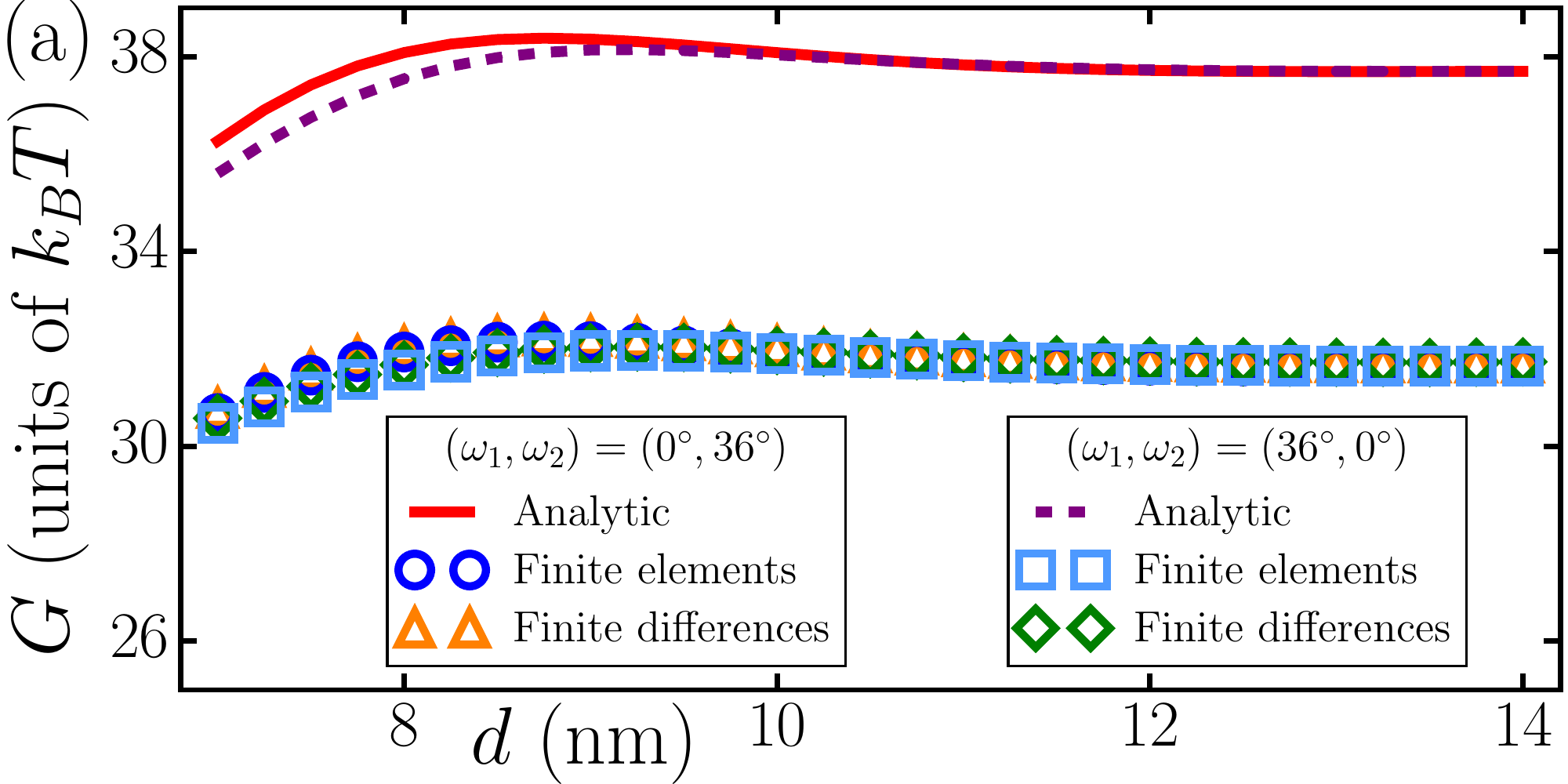}\\
\includegraphics[width=\columnwidth]{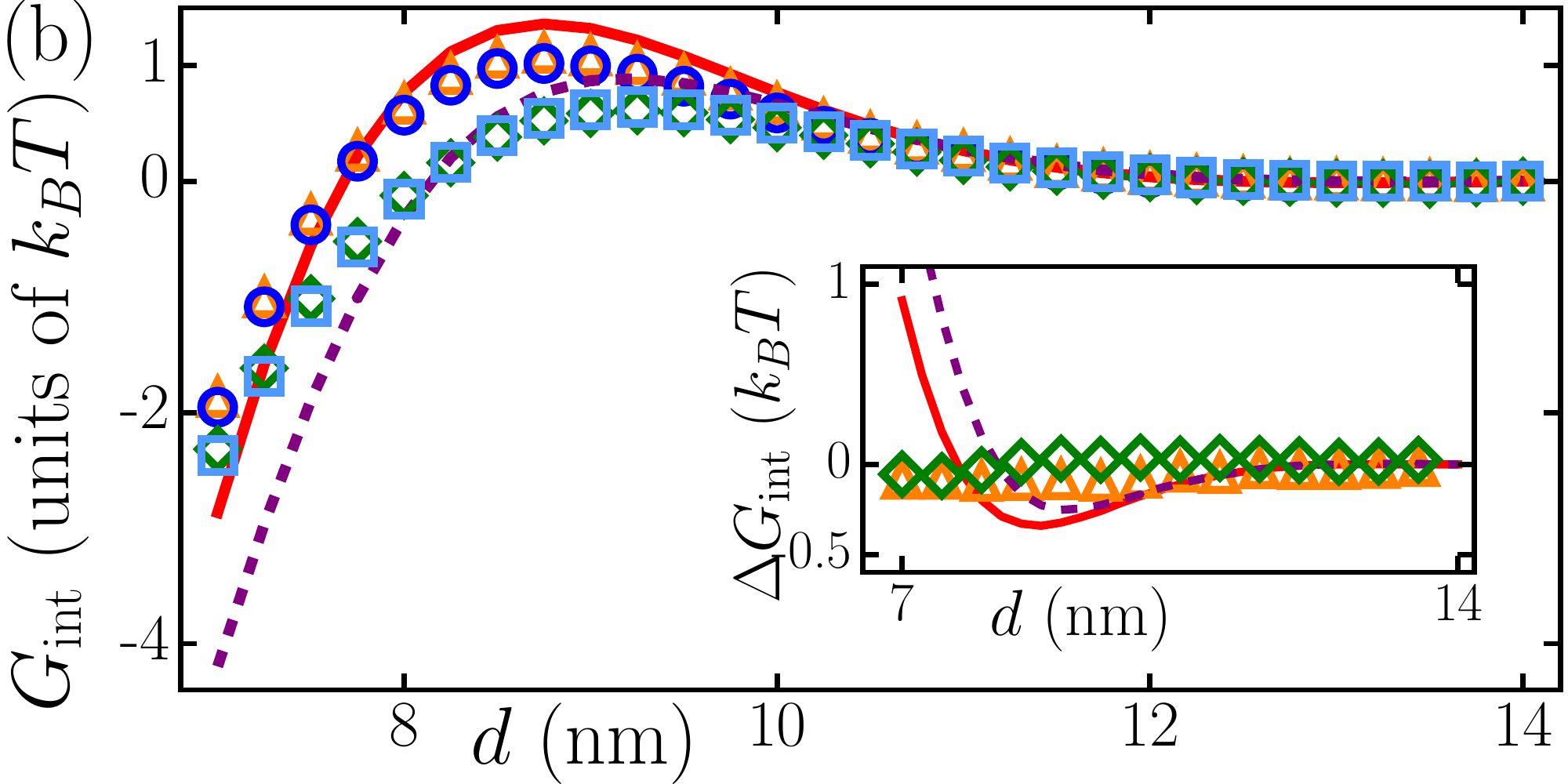}
\caption{(Color online) Thickness deformation energy of clover-leaf shapes. (a) Thickness deformation energy per protein for two clover-leaf shapes, $G$, versus center-to-center protein distance, $d$, calculated analytically at $N=11$, and numerically using FE and FD schemes, for the face-on configuration
[$\omega_1=0^\circ$ and $\omega_2=36^\circ$ in Eq.~(\ref{boundCclover})]
and the tip-on configuration [$\omega_1=36^\circ$ and $\omega_2=0^\circ$ in Eq.~(\ref{boundCclover})]. (b) Thickness-mediated interaction energy $G_\textrm{int}$ versus $d$ obtained by subtracting the protein-induced thickness deformation
energies in the non-interacting regime from the respective perturbative analytic, FE, and FD solutions in panel (a). The inset shows the difference in the thickness-mediated interaction energies obtained from the FE solution procedure, and the analytic and FD solution procedures. We use the same labeling conventions for panel (b) as for panel (a). For our numerical calculations we used $\langle l_\textrm{edge} \rangle \approx 0.26$~nm for the FE and $h=0.05$~nm for the FD solutions. We set $\tau=0$ for both panels. All model parameters were chosen as described in Sec.~\ref{secElasticModel}, and analytic and numerical solutions were obtained as discussed in Secs.~\ref{secAnalyticSol}
and~\ref{secNumericalSol}.
}
 \label{fig:clover5} 
\end{figure}

To compare our analytic and numerical solutions in the case of clover-leaf shapes, we consider the bilayer thickness deformation energies associated with two clover-leaf proteins in the ``face-on configuration'' [$\omega_1=0^\circ$ and $\omega_2=36^\circ$ in Eq.~(\ref{boundCclover})] and the ``tip-on configuration'' [$\omega_1=36^\circ$ and $\omega_2=0^\circ$ in Eq.~(\ref{boundCclover})] using analytic, FE, and FD solution procedures [see Fig.~\ref{fig:clover5}(a)]. We find that the thickness deformation energies obtained using our FE and FD solution procedures are in agreement within the numerical accuracy of the numerical solution schemes. In contrast, the thickness deformation energies obtained via the perturbative analytic solution procedure differ substantially from the FE and FD solutions, in the non-interacting as well as interacting regimes in Fig.~\ref{fig:clover5}(a). Some discrepancy between analytic and numerical results is to be expected, given that somewhat different boundary value problems are solved in the perturbative analytic and numerical approaches, with the analytic solution only being first order in the perturbation parameter $\epsilon_i$ in Eq.~(\ref{boundCclover}). This produces a systematic error in the thickness deformation energy obtained through the perturbative analytic approach, which yields disagreement between perturbative analytic and numerical solution procedures even in the non-interacting regime.

As far as bilayer thickness-mediated interactions between clover-leaf shapes are concerned [see Fig.~\ref{fig:clover5}(b)], the discrepancy between analytic and numerical results is most pronounced at small $d$, where the interaction potentials obtained from perturbative analytic and numerical approaches can differ by $>1$~$k_B T$. This can be understood intuitively by noting that the analytic calculation of thickness-mediated interactions between clover-leaf shapes relies on a perturbative mapping of clover-leaf boundary curves with
constant hydrophobic thickness onto circular boundary curves with varying hydrophobic thickness [see Eqs.~(\ref{genBC1trimerFin})--(\ref{genBC2trimerFin2})].
We use here a first-order perturbative mapping, which is expected to become increasingly inaccurate at small protein separations since, as $d$ is being
decreased, the structure of thickness deformations in close proximity to the clover-leaf proteins comes to dominate the interaction energy. We therefore attribute the disagreement between analytic and numerical approaches in Fig.~\ref{fig:clover5}(b) to shortcomings of the analytic approach. Note, however, that the first-order perturbative analytic solution and the corresponding numerical solutions yield the same basic scenario for the competition between the different orientations of pentameric clover-leaf shapes considered in Fig.~\ref{fig:clover5}. Indeed, similar agreement between first-order perturbative and numerical approaches is obtained for other clover-leaf shapes and orientations \cite{CAH2013a,OWC1}, which suggests that the first-order perturbative approach can accurately capture the directionality of thickness-mediated interactions between integral
membrane proteins with clover-leaf shapes.

\section{Summary and conclusions}
\label{secSummary}

A wide range of experiments indicate 
\cite{mouritsen93,jensen04,engelman05,andersen07,mcintosh06,phillips09,brown12,lundbaek06,harroun99,goforth03,botelho06,grage11}
that protein-induced lipid bilayer thickness deformations can play a crucial
role in the regulation of protein function through bilayer material properties and bilayer-mediated protein interactions.
Cell membranes are generally crowded with membrane proteins \cite{engelman05,takamori06,dupuy08,phillips09,linden12},
suggesting that the protein center-to-center distance $d$ is typically small
\textit{in vivo}, while modern structural biology suggests a rich picture of membrane protein shape with great diversity in the oligomeric states and symmetries of membrane proteins. Motivated by these experimental observations,
we have developed a combined analytic and numerical framework \cite{CAH2013a,CAH2013b,OWC1,OPWC1,CAH2014a} which allows prediction of the protein-induced lipid bilayer thickness deformations
implied by the classic model in Eq.~(\ref{energy}) for arbitrary $d$ and the protein shapes suggested by structural studies. Our analytic solution procedure, which is based on Refs.~\cite{huang86,dan93,dan94,aranda96,dan98,goulian93,weikl98}, allows exact solutions of the protein-induced lipid bilayer thickness deformation field and elastic thickness deformation energy for proteins with constant or varying hydrophobic thickness in the (strongly) interacting as well as non-interacting regimes, provided that the proteins have circular transmembrane
cross sections. Through a perturbative approach, our analytic solution scheme can also account for non-circular protein cross sections. Following similar steps as in Refs.~\cite{goulian93,weikl98,CAH2013a}, our analytic solution procedure is readily applied \cite{CAH2014a} to calculate curvature-mediated protein interactions at arbitrary protein separations.

The exact analytic solutions described here are in excellent quantitative agreement with our numerical solutions for arbitrary protein orientations and arbitrary $d$ with the exception of the strongly unfavorable regime of bilayer thickness-mediated interactions, for which the interaction energy diverges as the edge-to-edge protein separation approaches zero. We regard this regime as being of limited physical significance because, for proteins of distinct
hydrophobic thickness,  the leading-order model in Eq.~(\ref{energy}) is expected to break down at small $d$ due to the large gradients of protein-induced
lipid bilayer deformations in the bilayer region separating the two proteins. In principle,
higher-order analytic solutions than those considered here could be developed to access thickness-mediated interactions in this regime. For proteins
of non-circular cross section, our comparisons between analytic and numerical
solution procedures show that the first-order perturbative solution can accurately capture the dependence of the thickness deformation
energy on protein oligomeric state \cite{CAH2013b,OWC1}, as well as the directionality of bilayer thickness-mediated protein interactions \cite{CAH2013a,OWC1,OPWC1}. However, the first-order perturbative approach does not yield the exact value of the thickness deformation energy. The discrepancy between our perturbative analytic and numerical solutions of bilayer thickness-mediated protein interactions is most pronounced at small $d$, where the perturbative analytic approach is expected to break down.

We have developed both FD and FE solution schemes to numerically calculate the lipid bilayer thickness deformations induced by membrane proteins. The FD scheme has the advantage of being conceptually simple and relatively straightforward to implement. We find that our FD scheme accurately accounts
for the lipid bilayer thickness deformations induced by cylindrical membrane proteins in the non-interacting regime and can also
capture, albeit with less accuracy, bilayer thickness-mediated interactions between cylindrical membrane
proteins. However, the convergence of the numerical solutions to the exact analytic solutions is slower for the FD scheme than for the FE scheme, most likely due to errors in the FD slope boundary conditions at the protein surface. Furthermore, we find that the FD solution procedure can introduce substantial numerical errors for non-cylindrical membrane proteins. These errors are particularly pronounced in the interacting regime.

In contrast, we find that the FE scheme described here yields rapid numerical convergence for all available exact analytic solutions of the minima of Eq.~(\ref{energy})
\cite{huang86,CAH2013a,CAH2013b}. The convergence
properties of the FE scheme do not seem to be diminished substantially in
the interacting regime compared to the non-interacting regime of bilayer
thickness-mediated protein interactions, or by complicated boundary shapes and boundary conditions. The combined presence of both first and second derivatives in the energy in Eq.~(\ref{energy}) places special demands on the FE formulation. In particular, while standard Lagrange interpolation functions \cite{shames1985energy} are adequate to compute the thickness stretch and gradient terms in Eq.~(\ref{energy}), they fail to produce conforming curvatures at element interfaces \cite{bathe2006}. Our FE solution procedure overcomes this challenge by combining \cite{OWC1,OPWC1} Lagrange shape functions for the thickness stretch and gradient terms with
the DKT method \cite{Batoz1980,Bathe1981} for curvature deformations. The resulting FE approach is computationally efficient and allows accurate
solutions of the complicated boundary value problems posed by many
interacting membrane proteins. Indeed, we have shown previously \cite{OWC1,OPWC1}
that our FE approach permits calculation of directional thickness-mediated
protein interactions in systems composed of hundreds of integral membrane proteins for arbitrary protein separations and orientations using protein shapes suggested by membrane structural biology.

The combined analytic and numerical framework described here shows that the shape of integral membrane proteins, and resulting structure of lipid bilayer thickness deformations, can play a crucial role in the regulation of protein function by lipid bilayers \cite{OWC1,CAH2013b}, and that bilayer thickness-mediated interactions between integral membrane proteins are strongly directional and dependent on protein shape \cite{CAH2013a,OWC1,OPWC1,CAH2014a}. Taken together, our results suggest that, in addition to bilayer-protein hydrophobic mismatch
\cite{engelman05,mouritsen93,jensen04,mcintosh06,lundbaek06,andersen07,phillips09,brown12,harroun99,goforth03,botelho06,grage11},
protein shape may be a crucial determinant of membrane protein regulation
by lipid bilayers and bilayer-mediated protein interactions.

The classic model of protein-induced lipid bilayer thickness deformations in Eq.~(\ref{energy}) and modifications thereof have been found to capture the basic experimental
phenomenology of bilayer-protein interactions in a wide range of experimental systems
\cite{dan93,dan94,aranda96,dan98,harroun99b,partenskii04,brannigan07,ursell07,huang86,helfrich90,nielsen98,nielsen00,harroun99,partenskii02,partenskii03,kim12,lundbaek10,greisen11,wiggins04,wiggins05,ursell08,grage11,CAH2013a,CAH2013b,OWC1,OPWC1,mondal11,mondal12,mondal13,mondal14,CAH2014a,andersen07,phillips09,jensen04,lundbaek06,mcintosh06,brown12},
only involve parameters which can be measured directly in experiments, and are simple enough to allow analytic solutions. Analogous models have been
formulated \cite{fournier99,phillips09} to describe protein-induced curvature deformations 
\cite{goulian93,weikl98,kim98,kim00,muller05,muller05b,kim08,auth09,muller10,frese08,reynwar11,bahrami14,dommersnes99,evans03,weitz13,yolcu14}
and fluctuation-mediated interactions \cite{goulian93,dommersnes99,evans03,weitz13,yolcu14,golestanian96,golestanian1996b,weikl01,lin11}.
In general, thickness-, curvature-, and fluctuation-mediated interactions all contribute to bilayer-mediated interactions between integral membrane proteins, but the relative strengths of these interactions depend on the specific experimental system under consideration. In addition to bilayer-mediated
interactions, membrane proteins may, in principle, also interact via electrostatic forces. However, electrostatic interactions in aqueous environments are generally screened \cite{bental96,walther96} and charged protein residues are typically
excluded from the transmembrane regions of membrane proteins \cite{ulmschneider01}.

The model of protein-induced lipid bilayer thickness deformations in Eq.~(\ref{energy}) and the corresponding ``zeroth-order'' models
\cite{goulian93,weikl98,kim98,kim00,muller05,muller05b,kim08,auth09,muller10,frese08,reynwar11,bahrami14,dommersnes99,evans03,weitz13,yolcu14,golestanian96,golestanian1996b,weikl01,lin11}
capturing curvature- and fluctuation-mediated protein interactions absorb the molecular details of lipids and membrane proteins into effective material
parameters. To provide a more detailed description of bilayer-protein interactions, a number of extensions and refinements of these models have been developed
\cite{gil98,brannigan06,brannigan07,west09,may04,may99,bohinc03,may07,watson11,watson13,jablin14,rangamani14,bitbol12,partenskii02,partenskii03,partenskii04,kim12,yoo13,yoo13b,lee13}.
For instance, the effect of bilayer-protein interactions on the elastic properties of lipid bilayers can be captured \cite{partenskii02,partenskii03,partenskii04,lee13}
by allowing for spatial variations in the values of the elastic bilayer parameters. Furthermore, the microscopic roughness of lipid bilayers due to area fluctuations
can affect \cite{brannigan06,west09} protein-induced lipid bilayer deformations. These model refinements yield bilayer thickness deformation profiles which are quantitatively but not qualitatively different from those implied by Eq.~(\ref{energy}) [see, for instance, Fig.~\ref{fig:single_cyl}(a)].

However, additional structural properties of the lipid bilayer, such as lipid tilt \cite{may04,may99,bohinc03,may07,watson11,watson13, jablin14}, may have a more substantial effect on bilayer-mediated protein interactions. Moreover, integral membrane proteins may tilt to reduce hydrophobic mismatch, as suggested by Monte Carlo and molecular dynamics simulations \cite{kim10,neder12}. Tilting of membrane proteins is expected to be most pronounced for small membrane proteins with only a single transmembrane $\alpha$-helix. While protein tilting generally competes with protein-induced lipid bilayer thickness deformations as a mechanism for alleviating bilayer-protein hydrophobic mismatch, experiments have suggested \cite{dePlanque03,holt10} that protein tilting is in general too weak to fully offset a hydrophobic mismatch between membrane proteins and the surrounding lipid bilayer.

In this article we have focused on bilayer thickness-mediated interactions between two integral membrane proteins. In general, more than two membrane proteins are expected to interact in the crowded membrane environments provided by living cells. We have shown previously \cite{OPWC1} that, in contrast to curvature- and fluctuation-mediated protein interactions \cite{kim98,kim00,dommersnes99,kim08,weitz13,yolcu14}, bilayer thickness-mediated protein interactions are approximately pairwise additive, at least
for large enough protein separations. For small protein separations, non-pairwise contributions to bilayer thickness-mediated interactions between integral membrane proteins can modify the interaction strength by $>1k_B T$ \cite{OPWC1}. However, except in special cases \cite{OPWC1}, non-pairwise contributions to bilayer
thickness-mediated protein interactions do not alter how bilayer thickness-mediated interactions vary with the shape and arrangement of proteins. The approximate pairwise additivity of bilayer thickness-mediated protein interactions presents a considerable simplification \cite{OPWC1} for the mathematical analysis of systems composed of many interacting integral membrane proteins.

Recent breakthroughs in superresolution light microscopy and electron cryo-tomography have revealed that integral membrane proteins can form large clusters with intricate translational and orientational protein ordering, which provides
\cite{mouritsen93,jensen04,engelman05,andersen07,mcintosh06,phillips09,brown12,lundbaek06,bray04,lang10,anishkin13,anishkin14}
a general mechanism for cells to modulate protein function through cooperative interactions and local modification of bilayer mechanical properties. But,
to date, the physical mechanisms giving rise to the observed lattice architectures and collective functions of membrane protein clusters remain largely unknown.
The directionality of bilayer thickness-mediated protein interactions implied by the observed protein structures presents one possible physical mechanism for membrane protein organization and collective function. Such directional interactions can yield \cite{OWC1,OPWC1,CAH2013a,CAH2014a} ordering of integral membrane proteins, which is also consistent with molecular dynamics simulations \cite{periole07,parton11,mondal13,yoo13}. The combined analytic and numerical framework we have discussed here allows calculation of the lipid bilayer-mediated protein interactions implied by bilayer elasticity theory 
\cite{seifert97,boal02,safran03,canham70,helfrich73,evans74,huang86,dan93,dan94,aranda96,dan98,harroun99b} for the protein shapes suggested by structural studies at arbitrary protein separations and orientations. Our framework thus presents a step towards a general physical theory of how directional bilayer-mediated protein interactions affect the molecular structure, organization, and biological function of proteins in the crowded membrane environments provided by living cells.

\acknowledgments{This work was supported at USC by NSF Award No. DMR-1206332,
an Alfred P. Sloan Research Fellowship in Physics (C.A.H.), the James H. Zumberge Faculty Research and Innovation Fund at the University of Southern California, and by the USC Center for High-Performance Computing, and at UCLA by NSF Award No. CMMI-0748034 and No. DMR-1309423. We thank M. Lind\'en, R. Phillips, and N.~S. Wingreen for helpful comments.}

% \bibliography{references}

\end{document}